\documentclass[soumission]{jsfds}
\usepackage{hyperref}
\usepackage{amsmath,amsthm,amssymb,amsbsy,amsfonts} 
\usepackage{graphicx}
\usepackage{natbib}
\usepackage{verbatim}
\usepackage{times} 
\usepackage{geometry}
\usepackage{dsfont}
\usepackage{underscore}
\usepackage{color}
\usepackage{bm}
\usepackage{url}
\usepackage[titletoc,toc,title]{appendix}

\makeatletter
\newcommand{\verbatimfont}[1]{\def\verbatim@font{#1}}%
\makeatother

\usepackage{hyperref}
\hypersetup{pdfborder={0 0 0},
	colorlinks=true,
	linkcolor=blue,
	citecolor=red,
	urlcolor=blue,
	hyperindex=true}
\setmainlanguageenglish
\firstpage{1}
\begin{document}
\begin{frontmatter}

\title{Latent Gaussian modeling and INLA:
  \\ A review with focus on space-time applications}
\runtitle{INLA for space-time statistics}
\alttitle{Mod\`eles \`a processus gaussiens latents et inf\'erence
  INLA: \\  un
  survol orient\'e vers les applications spatio-temporelles}

\begin{aug}
	\author{
		\prenom{Thomas}
		\nom{Opitz}
		\thanksref{t1}
		\contact[label=e2]{thomas.opitz@inra.fr}
	}
	\affiliation[t1]{BioSP, INRA, F-84914 Avignon, France. \\
		thomas.opitz@inra.fr\\ 
	Please cite this review paper as \\ \emph{T. Opitz. Latent Gaussian modeling and INLA:
			A review with focus on space-time applications. To appear in the Journal of the French Statistical Society, 2017.}}
	\runauthor{T. Opitz}
\end{aug}

\begin{abstract}
Bayesian hierarchical models with latent Gaussian layers  have proven very flexible in capturing complex
stochastic behavior and hierarchical structures in high-dimensional spatial and spatio-temporal
data. Whereas simulation-based Bayesian inference through Markov Chain
Monte Carlo may be hampered by
slow convergence and numerical instabilities, the inferential framework of Integrated
Nested Laplace Approximation (INLA) is capable to provide accurate and relatively fast analytical
approximations to posterior quantities of interest.  It heavily relies on the use of
Gauss--Markov dependence structures to avoid the numerical bottleneck
of high-dimensional nonsparse matrix computations. With a view
towards space-time applications, we here review the principal
theoretical concepts, 
model classes  and inference tools within the
INLA framework. Important elements to construct space-time models are certain spatial Mat\'ern-like
Gauss--Markov random fields, obtained as approximate solutions to a
stochastic partial differential equation. Efficient implementation of statistical
inference tools for a large variety of models is available  through
the \texttt{INLA} package of the \texttt{R} software. 
To showcase the  practical use of \texttt{R-INLA} and to illustrate its principal commands and syntax,  a comprehensive simulation experiment is  presented using  simulated
non Gaussian space-time count data with a first-order autoregressive dependence
structure in time.
\end{abstract}

\begin{altabstract}
Les modè\`eles bay\'esiens hi\'erarchiques structur\'es par un processus
gaussien latent sont largement utilis\'es dans la pratique statistique
pour  caract\'eriser
des comportements stochastiques complexes et des structures
hi\'erarchiques dans les donn\'ees en grande dimension,  souvent spatiales ou
spatio-temporelles. Si des m\'ethodes d'inf\'erence bay\'esienne de type
MCMC, bas\'ees sur la 
simulation de la loi a posteriori, sont souvent entrav\'ees par une covergence lente et des
instabilit\'es num\'eriques, l'approche inf\'erentielle par INLA
("Integrated Nested Laplace Approximation") utilise des
approximations analytiques, souvent tr\`es pr\'ecises et relativement rapides, afin de
calculer des quantit\'es li\'ees aux lois  a posteriori  d'int\'er\^et. Cette technique
s'appuie fortement sur des structures de d\'ependance de type
Gauss--Markov afin d'\'eviter des difficult\'es num\'eriques dans les
calculs matriciels en grande dimension. En mettant l'accent sur
les applications spatio-temporelles, nous discutons ici les
principales notions th\'eoriques, les classes de mod\`eles accessibles et les outils
d'inf\'erence dans le contexte d'INLA. Certains champs Markoviens Gaussiens, obtenus comme solution approximative  d'une \'equation
diff\'erentielle partielle stochastique,  sont la base de la mod\'elisation
spatio-temporelle. Pour illustrer l'utilisation pratique du logiciel
\texttt{R-INLA} et la syntaxe de ses commandes principales, un
sc\'enario de simulation-r\'eestimation est pr\'esent\'e en d\'etail, bas\'e
sur des donn\'ees simul\'ees, spatio-temporelles et non gaussiennes,  avec une
structure de d\'ependance autor\'egressive dans le temps. 
\end{altabstract}
\begin{altkeywords}
\kwd{Integrated Nested Laplace Approximation}
\kwd{R-INLA}
\kwd{statistique spatio-temporelle}
\end{altkeywords}
\end{frontmatter}

\section{Introduction}
The rapidly increasing availability  of massive sets of  georeferenced
data has spawned a strong demand for  suitable statistical modeling
approaches to handle large and complex data. Bayesian hierarchical
models have become a key tool for capturing and explaining
complex stochastic structures in spatial or spatio-temporal
processes. Many of these models are based on  latent Gaussian
processes, typically embedded in a parameter characterizing the central tendency of the
distribution assumed for the likelihood of the data,
and extend the Gaussian random field modeling brought forward by classical
geostatistics. Using a conditional independence assumption for the
data process with respect to the latent Gaussian layer makes inference
tractable in many cases. Typically, closed-form expressions for the likelihood are not available for these
complex models, and simulation-based inference through Markov chain
Monte Carlo (MCMC) has become a standard approach for many models. An
important alternative, superior to MCMC inference under certain aspects,  has been developed through the idea
of Integrated Nested Laplace Approximation, proposed  in the JRSS
discussion paper of \citet{Rue.Martino.Chopin.2009}. Many case studies have been conducted
through INLA in the meantime, with space-time applications to
global climate data \citep{Lindgren.al.2011},
epidemiology \citep{Bisanzio.al.2011}, disease mapping and spread
\citep{Schroedle.al.2011,Schroedle.al.2012}, forest fires
\citep{Serra.al.2014,Gabriel.al.2016}, air pollution risk mapping
\citep{Cameletti.al.2013}, fishing practices
\citep{CosandeyGodin.al.2014} or econometrics \citep{Gomez.al.2015a}. More generally, INLA has been
successfully applied  to generalized linear mixed
models \citep{Fong.Rue.Wakefield.2009}, log-Gaussian Cox processes
\citep{Illian.al.2012,Gomez.al.2015} and survival models \citep{Martino.al.2011}, amongst many other
application fields. The recent monograph of
\citet{Blangiardo.Cameletti.2015} reviews INLA in detail and gives
many practical examples. 
Instead of applying simulation techniques to produce a representative sample of the
posterior distribution, INLA uses analytic Laplace
approximation and efficient numerical integration schemes  to
achieve highly accurate analytical approximation of posterior
quantities of interest with relatively small computing times. In
particular, we get  approximations of univariate posterior
marginals of model hyperparameters and of the latent Gaussian variables. By
making use of latent Gauss--Markov dependence structures, models
remain tractable even in  scenarios that are very high-dimensional in terms of
observed data and latent Gaussian variables. 

The INLA-based inference procedures are implemented in the
\texttt{R}-package \texttt{INLA} (referred to as \texttt{R-INLA} in the
following) for a large
variety of models, defined through basic building blocks of three
categories:  the (univariate) likelihood specification of data, the latent Gaussian
model and prior distributions for hyperparameters. 
 Functionality of
\texttt{R-INLA} is continuously extended
\citep{Martins.al.2013,Lindgren.al.2015,Rue.al.2016}. This review and the code
examples refer to \texttt{R-INLA} version 0.0-1463562937. The \texttt{R-INLA} software
project is hosted on \url{http://www.r-inla.org/}, where one can find
lots of INLA-related resources, amongst them 
details on the specification of likelihoods, latent models and priors,
a discussion forum with very active participation of the members of
the INLA core team, tutorials and codes, an FAQ section, etc.

\section{Modeling and estimation with INLA}
\subsection{Latent Gaussian modeling}
The structured latent Gaussian regression models amenable to
INLA-based inference can be defined in terms of three layers:
hyperparameters, latent Gaussian field, likelihood model. The univariate
likelihood captures the marginal distribution of data and is often
chosen as an exponential family (Gaussian, gamma, exponential, Weibull, Cox, binomial, Poisson, negative
binomial, ...) similar to the framework of generalized linear models, where models like the exponential, Weibull or Cox ones
are available as survival models allowing for right- and left-side
censoring. The
mean (or some other parameter related to the central tendency) of the likelihood distribution is determined by the latent 
Gaussian predictor through a link function such that $\mathbb{E}(y\mid
\eta)=h^{-1}(\eta)$ in case of the mean, where $y$ is the observation, $\eta$ is a
Gaussian predictor and $h$ is an appropriately chosen link function. Hyperparameters can appear
in the likelihood as dispersion parameters like the variance of the
Gaussian distribution, the overdispersion parameter of the negative
binomial one or the shape parameter of the gamma one, or they can characterize the structure of the
latent Gaussian model, for instance through variances, spatial
correlation parameters or autoregression coefficients.
Formally, this hierarchical model can be written as 
\begin{align}\label{eq:model}
 \bm \theta &\sim \pi(\bm \theta) & \text{hyperparameters}\\
\bm x\mid \bm \theta & \sim \mathcal{N}(\bm 0, \bm Q(\bm
\theta)^{-1}) &\text{latent Gaussian field} \\
\bm y\mid \bm x, \bm \theta &\sim  \prod_i \pi(y_i\mid \eta_i(\bm x), \bm
\theta)  &\text{observations}
\end{align}
where $\bm Q(\bm
\theta)$ is the precision matrix (\emph{i.e.}, inverse covariance
matrix) of the latent Gaussian vector $\bm x$ and    $\bm \eta(\bm x)=\bm A\bm x$ with the so-called
\emph{observation matrix} $\bm A$ that maps the latent variable vector $\bm x$ to
the predictors $\eta_i=\eta_i(\bm x)$ associated to observations $y_i$. 
If $\bm y$ and $\bm x$ can be high-dimensional when using INLA, an
important limitation concerns the hyperparameter vector $\bm \theta\in\Theta$
whose dimension should be moderate in practice, say $< 10$ if these hyperparameters are estimated with the default settings of \texttt{R-INLA} (although \texttt{R-INLA} supports using a higher number of hyperparameters); this is due to
numerical integration that has to be carried out over the hyperparameter
space $\Theta$. Notice that the Gaussian likelihood is particular since,
conditional to the hyperparameters, the observations are still
Gaussian. In practice, the precision hyperparameter of the Gaussian
likelihood (\emph{i.e.}, the inverse of its variance)
can  correspond to a measurement error or a nugget effect, and we can
fix a very high value for the precision hyperparameter if we want the
model for data $\bm y$ to correspond exactly to the latent Gaussian
predictors $\bm \eta$. 
The dependence structure between observations $y_i$ is  captured
principally by the precision
matrix $\bm Q(\bm \theta)$ of the latent field $\bm x$.  In practice, it is strongly
recommended or even 
indispensable from the point of view of computation time and memory
requirements to choose
Gauss--Markov structures with sparse $\bm Q(\bm \theta)$ whenever
model dimension is high. 

The resulting joint posterior density of latent variables $\bm x$ and
hyperparameters $\bm \theta$ is
\begin{equation}\label{eq:postjoint}
\pi(\bm x,\bm \theta \mid \bm y) \propto \exp\left(-0.5\bm x'\bm Q(\bm
\theta)\bm x+\sum_i \log \pi(y_i\mid \eta_i,\bm\theta)+ \log \pi(\bm \theta)\right).
\end{equation}
This density over a high-dimensional space does usually not
characterize one of the standard multivariate families and is
therefore difficult to interpret and to manipulate.
In practice, the main interest lies in the marginal posteriors of
hyperparameters $\theta_j$, of
latent variables $x_i$ and of the resulting predictors $\eta_i$, where the latter
can be included into $\bm x$ for notational convenience. Calculation
of these univariate posterior densities requires integrating with respect 
to $\bm \theta$ and $\bm x$:
\begin{align}
\pi(\theta_j\mid \bm y) &= \int\int \pi(\bm x, \bm \theta \mid \bm y)
{\color{red} \mathrm{d}\bm x}\, \mathrm{d}\bm \theta_{-j} =\int\pi(\bm
\theta\mid \bm y ) \mathrm{d}\bm\theta_{-j}, \label{eq:posttheta}\\
\pi(x_i\mid \bm y) &= \int\int \pi(\bm x, \bm \theta \mid \bm y) {\color{red}\mathrm{d}\bm x_{-i}}\, \mathrm{d}\bm \theta  =  \int \pi(x_i \mid \bm \theta, \bm y)
\pi(\bm \theta\mid \bm y)\,\mathrm{d}\bm\theta. \label{eq:postx}
\end{align}
We notice that the use of astutely  designed numerical integration schemes with respect
to the moderately dimensioned hyperparameter space $\Theta$ can yield
satisfactorily accurate approximation of the outer integral.  On the other hand, calculating the inner integral with respect to
$\bm x$, often of very high dimension ($\approx$ $10^2$ to $10^8$),  is
intricate.

\subsection{Gauss--Markov models}
We say that a  random vector $\bm x\mid \bm \theta \sim
\mathcal{N}(\bm 0, \bm Q^{-1})$ is Gauss--Markov if the number of
nonnull entries of its $n\times n$ precision matrix $\bm Q=(q_{ij})_{1\leq i,j\leq n}$ is $\mathcal{O}(n)$. Such sparse precision matrices allow
efficient numerical computation of matrix operations like
$LR$-decomposition (with sparse factors $L$ and $R$), determinant calculation,
matrix-vector products, etc. For instance, complexity of matrix
inversion decreases from $\mathcal{O}(n^3)$ for matrices
without any structural constraints to around $\mathcal{O}(n^{3/2})$
for sparse matrices. 
Using Gauss--Markov structures
fundamentally shifts the dependence characterization from covariance
matrices $\bm Q^{-1}$ to precision
matrices $\bm Q$. Notice that the conditional expectation is easily expressed through the
regression 
$\mathbb{E}(x_i\mid \bm x_{-i})=-\sum_{j\not=i} (q_{ij}/q_{ii}) x_j$ 
where only a small number of the sum terms, also called the
neighborhood of $x_i$,  are non-zero
owing to the sparse structure of $\bm Q$. The conditional variance is
$\mathbb{V}(x_i\mid \bm x_{-i})=1/q_{ii}$. 
\texttt{R-INLA} uses fast and efficient algorithms for sparse matrix
calculations \citep{Rue.Held.2005}, already implemented in the
\texttt{GMRFLib} library. For efficient calculations, it is important to make the precision
matrix $\bm Q$ ``as diagonal as possible'' by reordering variables
to regroup nonzero elements as close as possible to the
diagonal. \texttt{R-INLA} has implemented several of those reordering
strategies; see \citet{Rue.Held.2005} for more details on reordering
algorithms. If certain Gauss--Markov models exist for spatially indexed graphs,
useful covariance functions defined over $\mathbb{R}^d$ and  leading to Gauss--Markov covariance
matrices are difficult to establish. An exception is the very flexible approximate
Gauss--Markov representation  of Mat\'ern-like covariances based on
certain stochastic partial differential equations (often referred to as the \emph{SPDE
  approach} in the literature), which is
also implemented in \texttt{R-INLA}; see
Section \ref{sec:spde} for more details. 

\subsection{INLA}
The fundamental idea of INLA consists in applying the
device of Laplace approximation to integrate out high-dimensional
latent components. This theoretical foundation is combined with
efficient algorithms and numerical tricks and approximations to ensure
a fast yet accurate approximation of posterior marginal densities of
interest like those of the latent field $\bm x$ (including the predictors $\eta_i$) in \eqref{eq:postx} or of hyperparameters $\theta_j$ in \eqref{eq:posttheta}. Since the details of methods implemented in the INLA approximation are quite technical, we here content ourselves with a
presentation of the main ideas and the related options available in \texttt{R-INLA}.

\subsubsection{The principle of Laplace approximation}
\label{sec:laplappr}
We first recall the principle of the Laplace approximation and its
calculation in practice. Typically, one seeks to evaluate an integral
$\int f(\bm x)\, \mathrm{d}\bm x$, where the positive  integrand
function $f$, here written as  $f(\bm x)
= \exp(kg(\bm x))$ with a scale variable $k\geq 1$, is defined over a high-dimensional space and is 
``well-behaved'' in the sense that  it satisfies some minimal regularity requirements, is unimodal and its shape
is not too far from gaussianity; for instance, requiring strict log-concavity of $f$ is useful, see \citet{Saumard.al.2014}.  Since the integral value is mainly
determined by the behavior around the mode of $g$, a second-order Taylor approximation of $g$ can be substituted for $g$ to
calculate an approximate value of the integral. Assuming that $\bm
x^\star$ is the unique global maximum of $g$, we get  $g(\bm x) \approx g(\bm
x^\star)+   0.5(\bm
x-\bm x^\star)'\bm H(g)(\bm x^\star) (\bm x - \bm x^\star)$ for values $\bm x$ close to
$\bm x^\star$ with the Hessian matrix $\bm H(g)(\bm x^\star)$. Notice that
$-\bm H(g)(\bm x^\star)$ is positive definite.  
An approximate value of the integral can be
calculated using the fact that a multivariate Gaussian density integrates to $1$.
The resulting following integral approximation in dimension $d$ is expected to become
more and more accurate
for  higher values of $k$, \emph{i.e.}, when the area below the integrand $\exp(kg(\bm x))$ becomes concentrated more and more closely around the mode:
\begin{align}
\int_{-\bm \infty}^{\bm \infty}  f(\bm x)\, \mathrm{d}\bm x&=\int_{-\bm \infty}^{\bm \infty} \exp(kg(\bm x))\, \mathrm{d}\bm x \label{eq:laplint}\\  & \overset{k\rightarrow
  \infty}{\sim} \int_{-\bm\infty}^{\bm \infty} \exp(kg(\bm x^\star) + 0.5k(\bm
x-\bm x^\star)'\bm H(g)(\bm x^\star) (\bm x - \bm x^\star))\, \mathrm{d}\bm x
\nonumber \\
& = \left(\frac{2\pi}{k}\right)^{d/2} \left|\bm H(g)(\bm x^\star)
\right|^{-1/2} \exp(k g(\bm x^\star));\label{eq:laplappr}
\end{align}
here $a\sim b$ means that $a=b(1+O(1/k))$  \citep{Tierney.Kadane.1986}. 
In statistical practice, $k$ may represent the number of
i.i.d. replications, each of which has density $\exp(g(\bm
x))$. Higher values of $k$ usually lead to better
approximation, and more detailed formal results on the quality of approximation have
been derived \citep{Tierney.Kadane.1986,Rue.Martino.Chopin.2009}. Many of the 
models commonly estimated with INLA have no 
structure of strictly i.i.d. replication, but the Laplace approximation remains
sufficiently accurate in most cases since there usually still is a structure of internal
replication; ideally, for each latent variable $x_{i_0}$ we have at least several observations $y_i$ which contain information about  $x_{i_0}$  (and which are conditionally independent with respect to $\bm x$ by construction of the model).

In the context of INLA, the following observation will be interesting and
useful. Fix $k=1$ in \eqref{eq:laplint}  and suppose that $f(\bm
x)=\exp(g(\bm x))=\pi(\bm x, \bm \theta)$, where $\pi(\bm x, \bm \theta)$ is the joint
probability density of a random vector $(\bm x, \bm \theta)$. Then, in \eqref{eq:laplappr},
the term $\exp(g(\bm x^\star))$ is the value of $\pi$ at its mode $\bm x^\star$
for fixed $\bm\theta$, whereas  $(2\pi)^{d/2} \left|\bm H(g)(\bm x^\star)
\right|^{-1/2}$ is $1/\pi_G(\bm x^\star\mid \bm\theta)$ with $\pi_G$ a Gaussian
approximation with mean vector $\bm x^\star$ to the conditional density of
$\bm x\mid \bm \theta$. In practice, we can determine the mean $\bm \mu^\star=\bm x^\star$ and the precision
matrix $\bm Q^\star=-\bm H(g)(\bm x^\star)$ of $\pi_G$ through an iterative
Newton--Raphson optimization. Starting from the joint posterior
\eqref{eq:postjoint} of our latent Gaussian model,  we set $g(\bm x) =-0.5\bm x'\bm Q(\bm
\theta)\bm x+\sum_i \log \pi(y_i \mid \eta_i,\theta)$. 
We further write $g_i(x_i)=\log \pi(y_i\mid
x_i,\bm\theta)$ and calculate its second-order Taylor expansion
$g_i(x_i)\approx g_i(\mu_i^{(0)})+b_ix_i-0.5c_ix_i^2$. 
Without loss of
generality, we here assume that the linear predictor $\bm \eta$
corresponds to the  latent Gaussian vector $\bm x$.
We start the
iterative optimization with initial values  $\bm Q^{(1)}=\bm
Q+\mathrm{diag}(\bm c)$ and $\bm \mu^{(1)}$, where $\bm Q^{(1)}\bm
\mu^{(1)} = \bm b$. We then iterate this procedure until convergence 
such that $\bm \mu^{(j)} \rightarrow \bm \mu^\star=\bm x^\star$ and $\bm
Q^{(j)}\rightarrow \bm Q^\star = \bm Q + \mathrm{diag}(\bm c^\star)$, $j=1,2,\ldots$, $j\rightarrow\infty$, where an
appropriate convergence criterion must be used. 
Notice that the conditional independence assumption of observations
$y_i$ with respect to 
$(\eta_i,\bm\theta)$ allows preserving  the sparse structure in $\bm Q^\star$. Moreover, a strictly log-concave likelihood function $x_i\mapsto \pi(y_i\mid x_i,\bm\theta)$ ensures $c_i>0$ such that  $\bm Q^{(j)}$ are valid precision matrices and local curvature information around $\mu_i^{(j)}$ can be used for constructing a useful Gaussian approximation.
It is further possible to impose linear constraints $\bm M\bm x = \bm
e$ onto $\bm x$ and $\bm x^\star$ with given matrix $\bm M$ and vector $\bm e$ by using
the approach of conditioning through kriging \citep{Rue.Martino.Chopin.2009}.

\subsubsection{Posterior marginal densities of hyperparameters}
\label{sec:posthyp}
To calculate 
\begin{equation}\label{eq:pithetaj}
\pi(\theta_j\mid \bm y) = \int\int \pi(\bm x, \bm \theta \mid \bm y)
{\color{red} \mathrm{d}\bm x}\, \mathrm{d}\bm \theta_{-j} =\int\pi(\bm
\theta\mid \bm y ) \mathrm{d}\bm\theta_{-j},
\end{equation}
we use the Laplace approximation of the inner integral $\int \pi(\bm x, \bm \theta \mid \bm y)
\mathrm{d}\bm x=\pi(\bm \theta\mid \bm y)$ as described in Section \ref{sec:laplappr} such that
the approximated density $\tilde{\pi}$ satisfies
\begin{equation}\label{eq:thetaappr}
\tilde{\pi}(\bm \theta\mid \bm y) \propto \frac{\pi(\bm x, \bm
  \theta, \bm y)}{\pi_G(\bm x\mid \bm \theta, \bm y)}\mid_{\bm x=\bm x^\star(\bm\theta)}
\end{equation}
with $\bm x^\star(\bm \theta)$ the mode of the joint density $\pi(\bm x, \bm\theta, \bm
y)$ for fixed $(\bm\theta,\bm y)$ and  a Gaussian density $\pi_G$ that
approximates $\pi(\bm x\mid \bm \theta, \bm y)$:
\begin{equation}\label{eq:gaussappr}
\pi_G(\bm x\mid \bm \theta, \bm y)=(2\pi)^{n/2}|\bm Q^\star(\bm\theta)|^{1/2}\exp\left(-0.5(\bm x-\bm x^\star(\bm
  \theta))'\bm Q^\star(\bm \theta)(\bm x - \bm x^\star(\bm \theta))\right).
\end{equation}
Notice that the Gaussian approximation $\pi_G$ is exact if the data likelihood
$\pi(y_i\mid \eta_i, \bm\theta)$ itself is Gaussian. 
An approximation of the posterior marginal of $\theta_j$ in \eqref{eq:pithetaj} is now
obtained through a numerical integration with a set of integration nodes
$\bm \theta_\ell$ chosen from a numerical exploration of the surface of the density
$\tilde{\pi}(\bm \theta_{-j},\theta_j\mid \bm y)$ (with $\theta_j$
held fixed). This yields 
\begin{equation}\label{eq:apprint}
\tilde{\pi}(\theta_j\mid \bm y) = \sum_{\ell=1}^L \omega_\ell
\tilde{\pi}(\bm \theta_\ell\mid \bm y)
\end{equation}
with weights $\omega_\ell$ (which are chosen to be equal in the
approaches implemented in \texttt{R-INLA}).
In \texttt{R-INLA}, $\bm \theta_\ell$ can either be chosen as a grid
around the mode of $\tilde{\pi}(\bm \theta\mid \bm y)$
(\texttt{int.strategy="grid"}, the most costly variant), or through
a simpler so-called \emph{complete composite design} which is
less costly when the dimension of $\bm \theta$ is relatively large
(\texttt{int.strategy="ccd"}, the default approach), or we may use
only  one integration node given as the mode value
(\texttt{int.strategy="eb"}, corresponding to the idea of an
\emph{empirical Bayes} approach).

\subsubsection{Posterior marginal densities of the latent Gaussian field}
For calculating the marginal density $\pi(x_i\mid \bm y)$ of a latent
variable $x_i$, we lean on representation  \eqref{eq:postx}. 
Numerical integration with respect to $\bm \theta$ can be done in
analogy to the procedure described in Section \ref{sec:posthyp}, and the
Laplace approximation \eqref{eq:thetaappr} allows approximating
$\pi(\bm \theta\mid\bm y)$. It thus remains to (approximately)
evaluate $\pi(x_i \mid \bm \theta, \bm y)$. A simple and  fast solution
would be to use the univariate Gaussian approximation resulting from
the multivariate Gaussian approximation \eqref{eq:gaussappr} whose 
mean value is $x_i^\star(\bm \theta)$ and whose variance can easily
and quickly be
calculated from a partial inversion of the precision $\bm Q^\star(\bm \theta)$ \citep{Rue.2005}
(\texttt{strategy="gaussian"} in \texttt{R-INLA}). However, this Gaussian
approximation often fails to capture skewness behavior  and can generate
nonnegligible bias in certain cases -- an important exception to this
issue being the case where the data likelihood is Gaussian. 
In the general case, using again a Laplace-like approximation
\begin{equation}\label{eq:laplacexi}
\frac{\pi(\bm x, \bm\theta,\bm y)}{\pi_G(\bm x_{-i}\mid x_i,\bm
  \theta,\bm y)}\mid_{\bm x_{-i}=\bm x_{-i}^\star(x_i,\bm \theta)}
\end{equation}
with mode $\bm x_{-i}^\star(x_i,\bm \theta)$ of $\pi(\bm x,
\bm\theta,\bm y)$ for fixed $(x_i,\bm \theta,\bm y)$ would be
preferable, but is relatively costly (\texttt{strategy="laplace"} in \texttt{R-INLA}). Instead, \citet{Rue.Martino.Chopin.2009}
propose a so-called \emph{simplified Laplace approximation} based on third-order Taylor
developments of numerator and denominator  in \eqref{eq:laplacexi}
that satisfactorily remedies location and skewness inaccuracies of the
Gaussian approximation (\texttt{strategy="simplified.laplace"} in
\texttt{R-INLA}, the default). Notice that the ``Nested'' in INLA
refers to this second Laplace-like approximation. 

\section{Space-time modeling approaches}
Modeling trends over space and time and spatio-temporal dependence in
repeated spatial observations is paramount to understanding the
dynamics of processes observed over space and time. 
We here review approaches to integrating the time component into the
latent Gaussian predictor $\eta_{st}$ that are suitable for high-dimensional space-time
inference with INLA. In principle, any space-time Gaussian process
$\eta_{st}$ could be used, but the requirement of a Gauss--Markov
structure for fast matrix calculations and the current scope of models implemented in
\texttt{R-INLA} impose some constraints. Flexible,
Mat\'ern-like  spatial Gauss--Markov
models  with continuous sample paths are available in \texttt{R-INLA}
and will be discussed in Section
\ref{sec:spde}. A generalization of such purely spatial models  to flexible nonseparable space-time dependence
structures is still pending, but \texttt{R-INLA} allows extending such 
spatial models to capture temporal dependence through autoregressive
structures, which will be discussed in the following and explored in
the examples of Section \ref{sec:application}. 
New approaches to generic nonseparable space-time Gauss--Markov
model classes and their implementation in \texttt{R-INLA}  are projected by members from
the scientific community working on the SPDE approach. 
Any other nonseparable covariance model would in
principle be amenable to INLA-based inference without difficulty as long
as it keeps  a Markovian structure when using a high-dimensional
latent model. 

We start by formulating a generic latent Gaussian space-time model for
the predictor $\eta_{st}$ that covers many of the
models that can be fitted with \texttt{R-INLA}.  We denote by
$\zeta_{stk}$, $k=1,\ldots, K$ given covariate data, which may depend on
space or time only. In some cases, one may obtain useful dynamical structures by constructing artificial covariates at time $t$ based on the observations at time $t-1$.  Notice that in cases where covariates are not
available except at the observation points $(s_i,t_i)$, $i=1,\ldots,n$ of
$y_i$, one may have to include them into the observation vector
$\bm y$ and formulate a latent Gaussian model for their
interpolation over space and time. This would give a model that achieves both interpolation of
covariate values and prediction of $\bm y$ at the same time. Here, we
consider 
\begin{equation}\label{eq:gaussST}
\eta_{st} = x_0+ \sum_{k=1}^K x_k \zeta_{stk} + \sum_{k=1}^L
f_k(\zeta_{stk}) +  x_t + x_s + x_{st}.
\end{equation}
The linear coefficients $x_0,\ldots,x_K$ are known as \emph{fixed effects},
whereas functions $f_k(\cdot)$ and the processes $x_t$, $x_s$ and
$x_{st}$ are referred to as \emph{random effects}. Notice that $x_t$
and $x_s$ are to be understood as marginal models that capture purely
spatial or temporal marginal effects. 
We now shortly present some typical examples of Gaussian prior models
that are commonly used
and available in \texttt{R-INLA}  for the intercept $x_0$, the linear covariate effects $x_k$, the
nonlinear covariate effects $f_k(\cdot)$, the marginal temporal
effect $x_t$, the marginal spatial effect $x_s$ and the space-time
effect $x_{st}$. 

A temporal effect $x_t$ could be modeled as an autoregressive
process or as a random walk, where autoregression coefficients and the
step variance of the random walk represent hyperparameters that could
be estimated. 
A nonparametric spatial effect $x_s$ could be modeled with a
Gauss--Markov Mat\'ern-like prior random field, the details of whose
construction are presented in
the following section \ref{sec:spde}. A simple extension to a space-time
effect $x_{st}$
is obtained from considering independent
replicates of the spatial field for each time point. An important
class of space-time models $x_{st}$ that allows for temporal
dependence and  preserves the
Gauss--Markov structure are the stationary first-order autoregressive models
\begin{equation}\label{eq:star}
x_{st} =  a x_{s,t-1} + \sqrt{1-a^2} \varepsilon_{st},
\end{equation}
where $a\in (-1,1]$ and $\varepsilon_{st}$ is a stationary spatial innovation
process, i.i.d. in time, typically chosen as the Gauss--Markov
Mat\'ern-like field. If the process starts in $t=1$ with $x_{s,1}=\varepsilon_{s,1}$, then its marginal distributions are stationary. We here allow for $a=1$ to include the purely spatial case; temporal independence arises for $a=0$.  This AR model is a \emph{group model} where spatial
groups $\varepsilon_{st}$, $t=1,2,\ldots$, are joined through an AR group model. 
In \texttt{R-INLA}, it is possible to work with more general group
models that define a type of dependence like ``autoregressive'',
``exchangeable'', ``random walk'' or ``besag'' between certain groups of latent
variables. When marginal variances tend to increase over time, an interesting alternative to the autoregressive model \eqref{eq:star}  may be to link spatial random fields through a random walk structure such that $x_{st} =  x_{s,t-1} + \varepsilon_{st}$; the variance of the innovation fields $\varepsilon_{st}$ then determines the marginal variance at instant $t$.  It is further possible to specify certain graph structures
among latent variables; we here refer to \url{www.r-inla.org} for full
details about the specification of a large variety of available latent models. 
Owing to issues of
identifiability and model complexity, usual only a subset of the terms
in \eqref{eq:gaussST} is used to construct the latent field  in practical applications.


\subsection{Spatial Gauss--Markov random fields based on the SPDE
  approach}
\label{sec:spde}
The spatial SPDE model of \citet{Lindgren.al.2011} defines
a Gauss--Markov random field as the approximate solution to a certain
stochastic partial differential equation. It is an important building
block for latent Gaussian models with spatial and spatio-temporal effects. Contrary to
classical covariance function models, this approach provides sparse precision
matrices that make numerical procedures efficient even for very
high-dimensional problems. 
Formally, a Gaussian process 
$x(s)$ on $\mathbb{R}^D$ is defined through
\begin{equation}
\label{eq:spde}
\left(\kappa^2 - \Delta \right)^{\alpha/2} x(s)  =W(s),
\quad  \alpha=\nu+D/2, \quad s\in \Omega
\end{equation}
with the Laplace operator $\Delta y=\sum_{j=1}^D \partial^2 y/\partial^2 x_j$,  a standard Gaussian white noise process $W(s)$ and a nonempty
spatial domain $\Omega\subset\mathbb{R}^D$ with regular boundary. Depending on the value
of $\nu$ and $D$, the Laplace operator $\left(\kappa^2 - \Delta
\right)^{\alpha/2}$ is fractionary with noninteger exponent $\alpha/2$,  and it must be defined in an
appropriate way \citep{Lindgren.al.2011}.
The only stationary solution to \eqref{eq:spde} for
$\Omega=\mathbb{R}^D$ is a Gaussian random
field with the Mat\'ern covariance function whose shape parameter is
$\nu$ (with  $\nu=0.5$ yielding the exponential covariance model) and whose scale parameter is $1/\kappa$.   The marginal
variance is $\Gamma(\nu)/[\Gamma(\nu+D/2)(4\pi)^{D/2}\kappa^{2\nu}]$,
and the ``empirical range'' where a correlation of approximately $0.1$ is attained between two
points is around $\sqrt{8\nu}/\kappa$.  The Mat\'ern model is known
to be very flexible through its scale and shape parametrization, with
regularity properties of sample paths governed by the shape parameter
$\nu$.

In practice, when working on a finite domain $\Omega\subset\mathbb{R}^2$, boundary effects come
into play. One can  assume 
a polygon-shaped boundary  $\partial\Omega$, as it is implemented in \texttt{R-INLA}.
An interesting choice of boundary
condition is the Neumann 
condition with  zero normal derivatives at the boundary 
such that the Gaussian field is ``reflected'' at the boundary.
\citet{Lindgren.al.2011} show that Neumann conditions
principally lead to an increase in variance close to the boundary, the
factor being approximately $2$ when there is one close linear boundary
segment, and $4$ when we are close to the $90$-degree angle of a rectangle where
two linear segments meet. Whereas such boundary conditions may be
interesting for some applications, we often prefer to extend the
domain $\Omega$ beyond the study region towards a larger domain, such that
boundary effects become negligible within the study region. This requires that the extended domain's boundary is
separated by a distance superior to  the empirical range from the
study region.

\begin{figure}
\centering
\includegraphics[width=10cm]{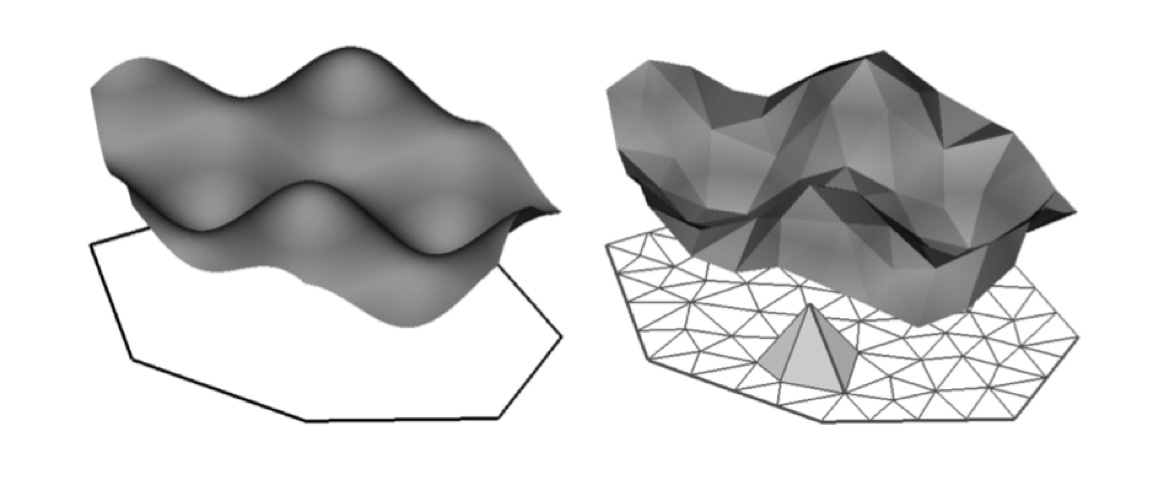}
\caption{Illustration taken
  from \citet{Blangiardo.al.2013}. Example of a Gaussian surface (left) and its finite element
  approximation (right). On the right display, we can further see the
  triangulation and one of the finite element pyramid-shaped basis functions. }
\label{fig:fembasis}
\end{figure}

Approximate solutions to \eqref{eq:spde} are obtained based on the
finite element approach commonly used in the numerical solution of
partial differential equations. Using a triangulation of the spatial domain leads to a high-dimensional
multivariate representation with Gaussian variables located in the
triangulation nodes $s_i$. Spatial interpolation between nodes is achieved  by considering these Gaussian
variables as weights for  piecewise linear basis functions $\psi_i(s)$,
one for each node. Basis functions are of compact
support giving by the triangles touching the node $s_i$ (``pyramid functions''); see Figure
\ref{fig:fembasis} for an example of a basis function and of the
approximation of a spatial surface through a linear combination of
such basis functions. By ``projecting'' the SPDE \eqref{eq:spde} on
the space spanned by the Gaussian variables, one can calculate the
precision matrix $\bm Q$ governing the dependence structure between
these variables. 
There are certain rules of thumb to be respected for a construction
of the triangulation that does not strongly distort the actual dependence
structure of the exact solution to \eqref{eq:spde} and that remains
numerically stable with respect to certain matrix computations, mainly concerning
maximum sizes of triangles and minimum sizes of interior angles. For
numerical efficiency, overly fine triangulations can be avoided by
requiring minimum edge lengths, for instance; see
\citet{Lindgren.al.2011} for further details on the finite
element construction. The approximate Gauss--Markov solution has mean
zero and its precision matrix $\bm Q$ has entries that are
determined by the SPDE. We refer to the Appendix section of \citet{Lindgren.al.2011} for the
calculation of $\bm Q$, which is explicit. Based on the approximate solutions for
$\alpha=0$ and $\alpha=1$, an approximate solution of the SPDE  for $\alpha+2$ can be obtained
by injecting the solution for $\alpha$ at the place of the white noise
$W(s)$ in \eqref{eq:spde}. For non-integer values of $\alpha>0$,
additional approximations are necessary to obtain a solution, see the authors' response in
the  discussion of the
\citet{Lindgren.al.2011} paper.  We have here considered  $\kappa$ and $\tau$ to
be constant over space. It is possible to  allow spatial
variation of these parameters for nonstationary models with
$\kappa=\kappa(s)$ as in \eqref{eq:spde} and a precision-related
parameter $\tau=\tau(s)$ that varies over space. 
To wit, \cite{Ingebrigtsen.al.2014} apply such second-order nonstationary
modeling to precipitation in
Norway, where altitude is used as a covariate that acts on the local covariance
range in the dependence structure; the recent contribution of \citet{Bakka.al.2016} uses second-order nonstationarity to account for physical barriers in species distribution modeling. The SPDE \eqref{eq:spde}  is also well-defined over manifolds $\Omega$, for instance the sphere
$\mathbb{S}_2$ embedded in $\mathbb{R}^3$. 
In general, \citet{Lindgren.al.2011} show that the approximate solution
converges to the true solution of the SPDE for an adequately chosen
norm when the triangulation is refined in a way such that the maximum
diameter of a circle inscribed into one of the triangles tends to
$0$. 

\texttt{R-INLA} currently implements calculation, estimation and simulation of the Gauss--Markov SPDE solution for $\alpha\leq 2$, for $\Omega$ a subset of
$\mathbb{R}^1$, 
$\mathbb{R}^2$ or a two-dimensional manifold embedded into $\mathbb{R}^3$, and for
spatially constant or varying values $\kappa(s)$ and $\tau(s)$. A
large number of tools is available to construct numerically efficient
and stable triangulations.

\section{Using \texttt{R-INLA} on a  complex simulated
  space-time data set}
\label{sec:application}
To execute the following code, the \texttt{INLA} package of \texttt{R} must be installed; see \url{www.r-inla.org}
for information on the \texttt{R} command recommended for installing
this package, which is not hosted on the site of the Comprehensive R
Archive Network (CRAN) due to its use of external libraries. We here
illustrate the powerful estimation and inference tools of
\texttt{R-INLA} in a controlled simulation experiment with data
simulated from a latent Gaussian space-time model with the Poisson
likelihood. The full code for the simulation study below can be requested from
the author. 

\subsection{Simulating the data}

We simulate a space-time count model based on a latent first-order
autoregressive Gaussian
process defined on $[0,1]^2$ for $t=1,\ldots,60$. We use two
covariates, given as $z_1(t)=t/60$ and $z_2(t)$ simulated according to an
autoregressive time series model. The simulated model  is
\begin{align*}
   Y(s,t)&\mid \eta(s,t) \sim \mathrm{Pois}(\exp(\eta(s,t))) \quad \text{i.i.d.} \\
  \eta(s,t) &=  -1+z_1(t)+0.5z_2(t) + W(s,t) \\
   W(s,1)&=\varepsilon(s,1) \\
W(s,t) &= 0.5W(s,t-1) +(1-0.5)^2 \varepsilon(s,t),\quad  t=2,\ldots,60.
\end{align*}
The covariance function of the standard Gaussian field
$\varepsilon(s,t)$ is chosen of Mat\'ern type with shape parameter $\nu=1$
and effective range $0.25$ (corresponding to a Mat\'ern scale parameter $1/\kappa\approx 0.09$). We now fix $50\times 60=3000$ observation points $(s_i,t_i)$ of
$Y(s,t)$, determined as the Cartesian product of $50$ sites uniformly
scattered in $[0,1]$ and $t_i=i$, $i=1,\ldots,60$. 

\begin{figure}
\centering
\includegraphics[height=4.5cm]{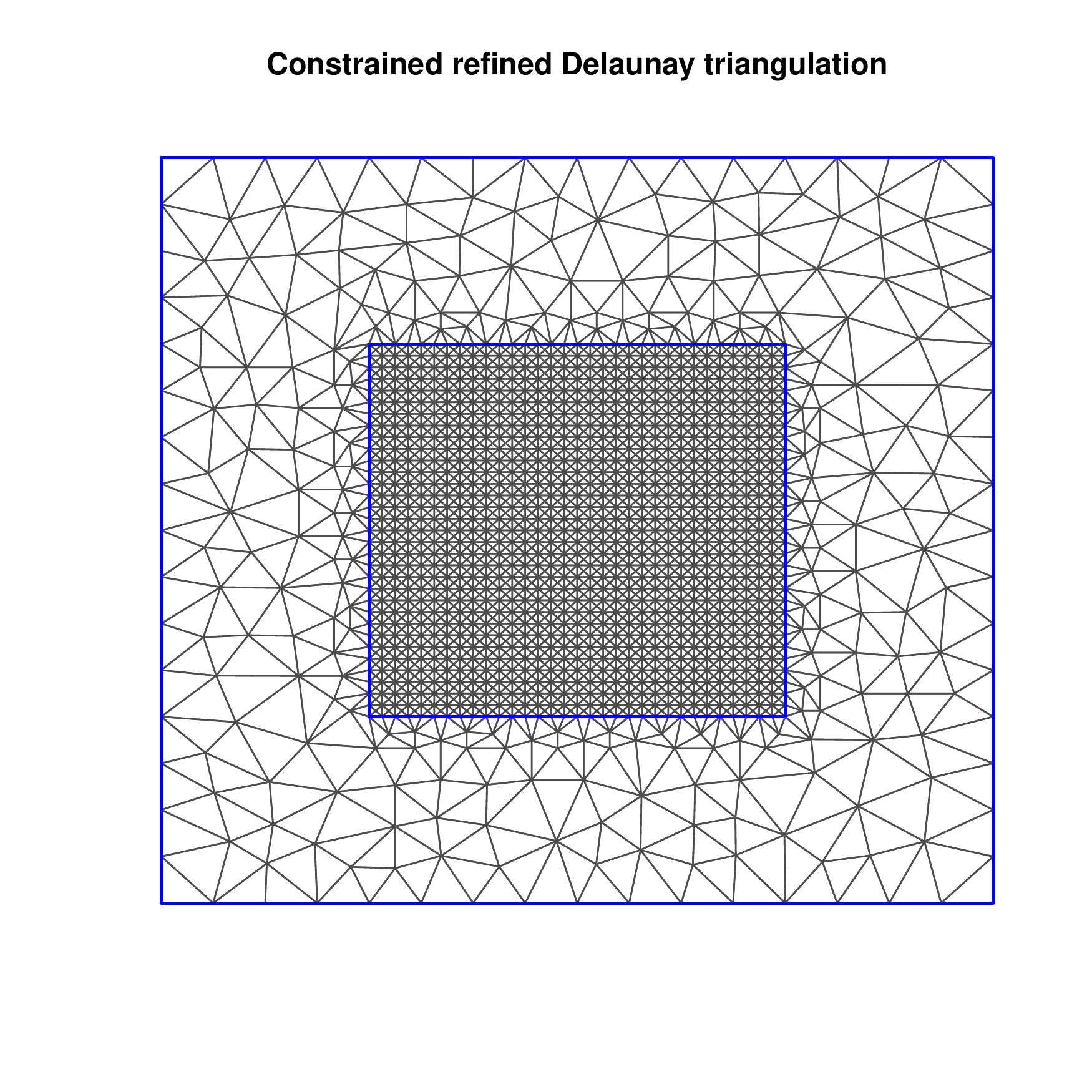}\quad
\includegraphics[height=4.5cm]{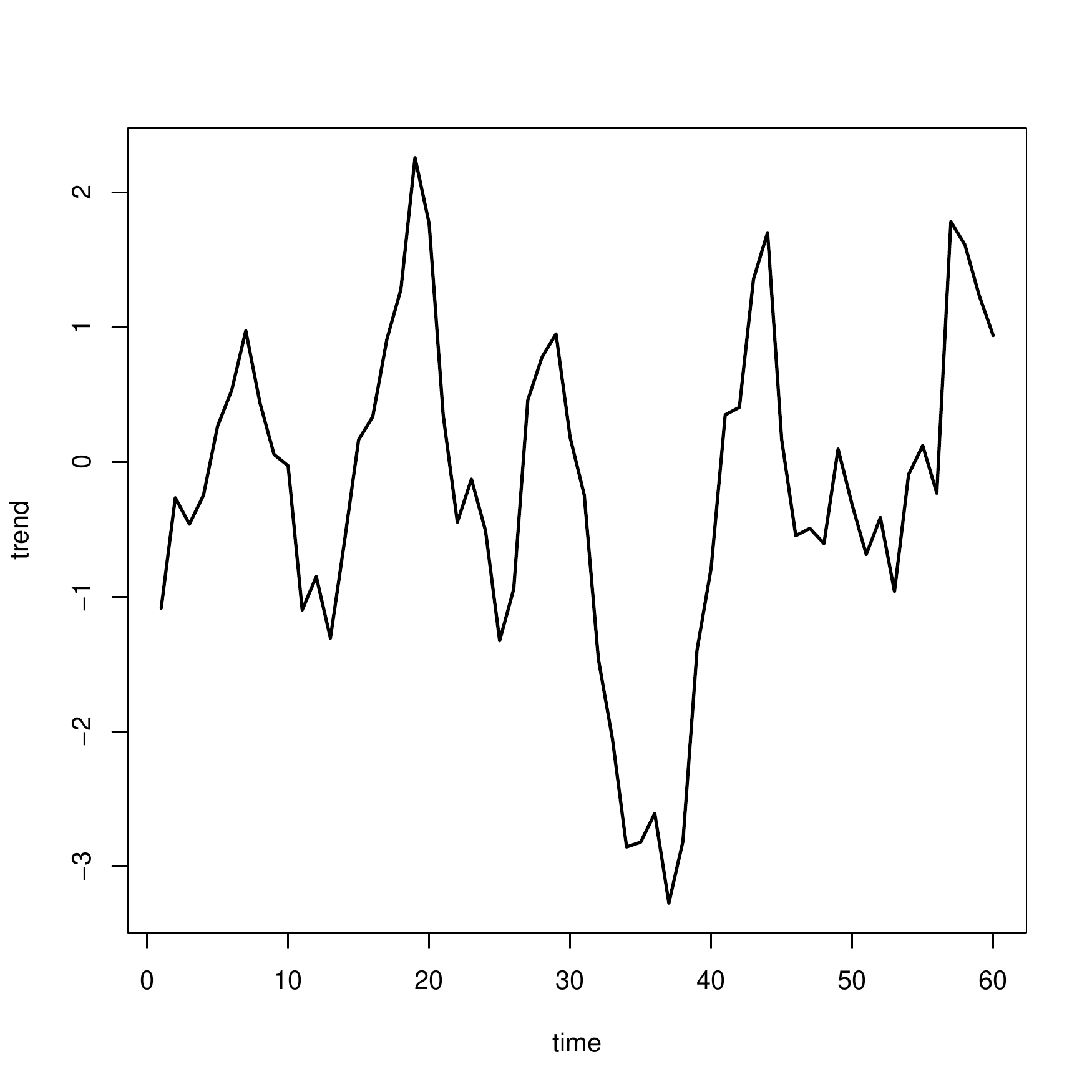}\quad
\includegraphics[height=4.5cm]{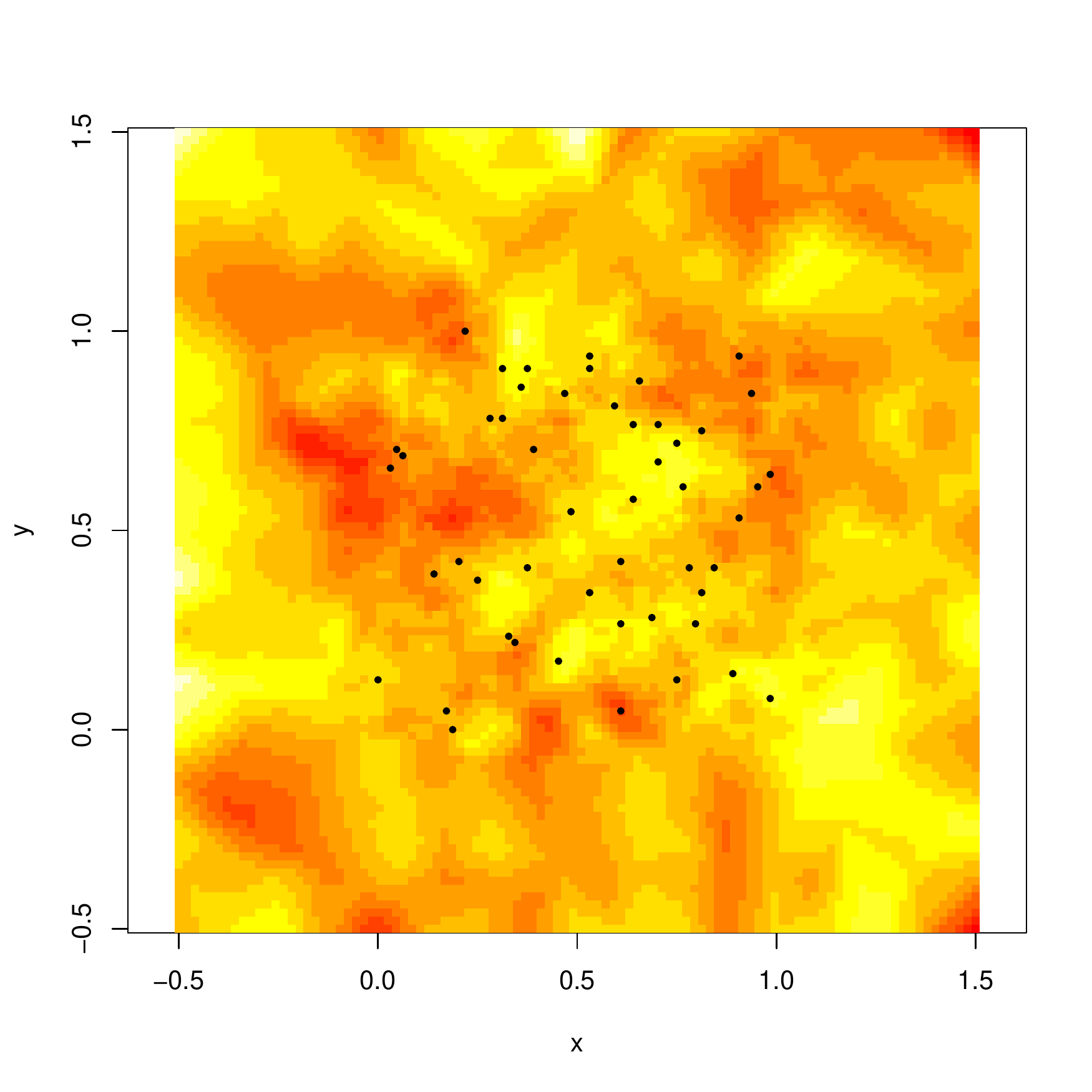}
\caption{Simulated latent Gaussian model. Left: triangulation used for
  the approximate SPDE simulation. Middle: time trend. Right:
  simulation of the linear predictor for $t=19$, with observation sites marked by black dots. }
\label{fig:gauss}
\end{figure}

To illustrate the simulation capacities of \texttt{R-INLA}, we here use
the SPDE approach to achieve simulation based on the Gauss--Markov  approximation of the
Mat\'ern correlation structure of the spatial innovations. 
After loading the \texttt{INLA}-package and fixing a random seed for better
reproducibility of results, we start by defining the $\kappa$,
$\tau$ and $\alpha$ parameters of the SPDE. To avoid boundary effects
in the SPDE simulation, we will use the square $[-0.5,1.5]^2$ as spatial
support. 
{\footnotesize
\begin{verbatim}
library(INLA)
seed=2;set.seed(seed)
n.repl=60;n.sites=50
nu=1;alpha=nu+1;range0=0.25;sigma0=1;a=.5
kappa=sqrt(8*nu)/range0
tau=1/(2*sqrt(pi)*kappa*sigma0)
\end{verbatim}
}
Next, we create a fine 2D triangulation mesh for relatively accurate simulation, with maximum
edge length $0.04$ within $[0,1]^2$ and $0.2$ in
$[-0.5,1.5]^2\setminus [0,1]^2$. The minimum angle between two edges
is set to $21$, a value recommended to avoid ill-conditioned
triangulations containing very elongated triangles.  Polygon nodes should be given in
counterclockwise order, whereas polygon-shaped holes in the support
would be specified by clockwise order of nodes; see the left display
of Figure \ref{fig:gauss} for the resulting triangulation.
{\footnotesize
\begin{verbatim}
nodes.bnd=matrix(c(0,0,1,0,1,1,0,1),ncol=2,byrow=T)
segm.bnd=inla.mesh.segment(nodes.bnd)
nodes.bnd.ext=matrix(c(-.5,-.5,1.5,-.5,1.5,1.5,-.5,1.5),ncol=2,byrow=T)
segm.bnd.ext=inla.mesh.segment(nodes.bnd.ext)
mesh.sim=inla.mesh.2d(boundary=list(segm.bnd,segm.bnd.ext),
   max.edge=c(.04,.2),min.angle=21)
plot(mesh.sim)
\end{verbatim}
}
The list slot \texttt{mesh.sim\$n} informs us that there are  $2401$
triangulation nodes.  The mesh object \texttt{mesh.sim} has a slot \texttt{mesh\$loc} which contains a three
column matrix. The first two columns indicate the 2D coordinates of
mesh nodes. In our case, the third coordinate, useful for specifying
2D manifolds in 3D space,  is constant $0$.
We continue by creating the SPDE model through an \texttt{R-INLA}
function \texttt{inla.spde2.matern}
whose main arguments are used to pass the mesh object and to fix
parameters $\alpha$, $\tau$ and $\kappa$. Its arguments \texttt{B.tau}
and \texttt{B.kappa} are matrices with one row for each mesh node.  If
only $1$ row is given, it describes a model with stationary values of
$\kappa$ and $\tau$, which will be 
duplicated internally for all nodes. For simulating a model with fixed
parameters $\kappa$ and $\tau$, these matrices have only one column that must be specified
as $\log \kappa$ or $\log \tau$ respectively. Then, we extract the
precision matrix $\bm Q$ of the resulting SPDE model and use it to create
independent samples of $\varepsilon(s,t)$, $t=1,\ldots,60$ through the
function \texttt{inla.qsample}. We fix the random seed for simulation
through its \texttt{seed=...} argument. Finally,
we manually create the first order AR model with coefficient $0.5$.
{\footnotesize
\begin{verbatim}
B.kappa=matrix(log(kappa),1,1)
B.tau=matrix(log(tau),1,1)
model.sim=inla.spde2.matern(mesh.sim,alpha=alpha,
   B.tau=B.tau,B.kappa=B.kappa)
Q=inla.spde2.precision(model.sim)
x=inla.qsample(n=n.repl,Q,seed=seed)
a=.5
for(i in 2:n.repl){x[,i]=a*x[,i-1]+sqrt(1-a^2)*x[,i]}
\end{verbatim}
}
It remains to fix covariate values and to generate the time trend in the
mean to add it to the centered Gauss--Markov space-time field. We fix an intercept $-1$  and the  two covariates \texttt{covar1}
\texttt{covar2}.
{\footnotesize
\begin{verbatim}
covar1=1:n.repl/n.repl
covar2=as.numeric(arima.sim(n=n.repl,model=list(ma=rep(1,5))))  
xtrend=-1+covar1+0.5*covar2
x=t(t(x)+xtrend)
plot(xtrend,type="l",xlab="time",ylab="trend",lwd=2)
\end{verbatim}
}
For the observed data $\bm y$ to be used in estimation, we sample
uniformly $50$ sites among the triangulation nodes contained in
$[0,1]^2$. By using \texttt{R-INLA}'s methods
\texttt{inla.mesh.projector} and \texttt{inla.mesh.project} to
project a spatial field with known values  for triangulation nodes
onto a regular grid necessary for standard plotting methods,  we
further provide a plot of $W(s,19)$ and the observation sites. At
$t=19$, we observe the maximum value of the time trend, and we will later
use \texttt{R-INLA} to do
spatial prediction for $t=19$.
{\footnotesize
\begin{verbatim}
nodes=mesh.sim$loc[,1:2]
idx.inside=which(pmin(nodes[,1],nodes[,2])>=0&pmax(nodes[,1],nodes[,2])<=1)
idx.obs=sample(idx.inside,size=n.sites)
sites=nodes[idx.obs,]
eta.i=as.numeric(x[idx.obs,])
y=rpois(length(eta.i),lambda=exp(eta.i))
t.pred=which.max(xtrend)
n.grid=100
grid=inla.mesh.projector(mesh.sim,dims=c(n.grid,n.grid))
image(grid$x,grid$y,inla.mesh.project(grid,field=x[,t.pred]),xlab="x",
   ylab="y",asp=1)
points(sites,pch=19, cex=.5)
\end{verbatim}
}
Figure \ref{fig:gauss} shows the trend component $-1+z_1(t)+0.5z_2(t)$ (middle
display) and the spatial field $\eta(s,19)$ at a fixed instant $t=19$ with observation sites
indicated (right display). 

\subsection{Fitting a complex space-time model with INLA}
\label{sec:fitmod1}
We now define and fit different candidate models to the above data
$\texttt{y}$ and its covariates \texttt{covar1} and \texttt{covar2}.  In Section \ref{sec:gof}, we
will then explore tools for  goodness-of-fit checks and model selection
within the \texttt{R-INLA} package.  We will first consider a model with prior structure similar to the simulated one, but we will also compare it to models where covariates are missing,
where the Mat{\'e}rn shape parameter $\nu$ takes a different value or
where the likelihood is not Poisson, but of the negative binomial type
(with an additional hyperparameter for overdispersion). 

First, let us define the triangulation mesh and the corresponding
prior spatial SPDE model for estimation. 
For estimation, we must be careful about the dimension of the latent
model to minimize memory requirements and high-dimensional matrix
calculations. Therefore, will use a mesh with a lower resolution than for
simulation, which may slightly increase the approximation error with
respect to the stationary Mat\'ern correlation structure.   It is often reasonable to use observation sites as initial nodes of the triangulation and to refine it by adding nodes
where necessary, or by removing nodes too close together which could be
source for numerical instabilitities. \texttt{R-INLA} implements
standard methods from the finite element literature and offers a 
conveniently parametrized interface to produce numerically stable and
moderately dimensioned triangulations. 
 \texttt{R-INLA}'s prior parametrization of
$\kappa$ and $\tau$ is a bit technical; it essentially assumes that  $\log(\tau)=b_{\tau,0} +
\theta_1 b_{\tau,1}+\theta_2 b_{\tau,2}+\ldots$ and $\log(\kappa)=b_{\kappa,0} +\theta_1 b_{\kappa,1}+\theta_2 b_{\kappa,2}+\ldots$, where the values  $b_i$ correspond to the 
columns of \texttt{B.tau} and \texttt{B.kappa}. The first column $b_0$ is a
fixed offset that must always be specified (even if it is $0$), whereas the following columns correspond to 
hyperparameters $\theta_i$ that are estimated when $b_i\not=0$ for  $\tau$
or $\kappa$. For instance, specifying \texttt{B.tau$=$B.kappa$=$matrix(c(0,1),nrow$=$1)} would lead to a model where  $\kappa=\tau=\exp(\theta_1)$.
In our model, we fix
the offset $b_0=0$ and we estimate two hyperparameters, one
corresponding to $\log(\tau)$, the other to $\log(\kappa)$.  This can
be seen as the standard prior specification of the SPDE model in
\texttt{R-INLA}. We fix the ``correct'' simulated value of
$\alpha$ in the SPDE model.
{\footnotesize
\begin{verbatim}
mesh=inla.mesh.2d(sites,offset=c(-0.125,-.25),cutoff=0.075,min.angle=21,
   max.edge=c(.1,.25))
plot(mesh)
points(sites,col="red",pch=19,cex=.5)
spde=inla.spde2.matern(mesh=mesh,alpha=alpha,B.tau=matrix(c(0,1,0),nrow=1),
   B.kappa=matrix(c(0,0,1),nrow=1))
\end{verbatim}
}
The mesh counts $294$ nodes. 
Notice that further arguments of the \texttt{inla.spde2.matern(...)}
function can be set to modify default priors, impose
"integrate-to-zero" constraints, etc. 
We now run a first estimation by considering a model with Poisson likelihood and prior
structure of the latent Gaussian field  corresponding to
the model that we simulated to generate the data. Naturally, this
model should provide a good fit  and we will later compare it to a
number of alternative models. 
The observation matrix $\bm A$ links observations $\bm
y$ to the latent variables $\bm x$ through $\bm A\bm x=\bm y$ and must
therefore be of dimension
 $ (50\times 60)\times$(\texttt{number of latent variables}) for our
 model. Since construction of this matrix and certain preprocessing
 steps before estimation like the removal of duplicate rows is rather complicated for
 complex models involving the spatial SPDE solution, \texttt{R-INLA} has helper
 methods that allow constructing this matrix and keeping track of
 latent variable indices more easily. In the following, the
 \texttt{inla.spde.make.index} command creates an index named
 "spatial", \emph{i.e.}, a data frame with a vector \texttt{spatial}
 (an index to match latent variables and triangulation nodes),
a vector \texttt{spatial.group} (an index that indicates the membership of a
 latent variable in a  "group" , here
 given as the instant $t\in\{1,\ldots,60\}$), and
a vector \texttt{spatial.repl} (an index that indicates the group membership
 when groups are  i.i.d.  replicates). In our case, all values of 
 \texttt{spatial.repl} are $1$ since there is no structure of
 i.i.d. blocks in 
 our space-time model due to the non-zero autoregression coefficient. 
{\footnotesize
\begin{verbatim}
idx.spatial=inla.spde.make.index("spatial",n.spde=spde$n.spde,n.group=n.repl)
A.obs=inla.spde.make.A(mesh,loc=sites,index=rep(1:nrow(sites),n.repl),
   group=rep(1:n.repl,each=nrow(sites)))
stack.obs=inla.stack(data=list(y=y),A=list(A.obs,1),effects=list(idx.spatial,
   data.frame(intercept=1,covar1=rep(covar1,each=n.sites),
      covar2=rep(covar2,each=n.sites))),tag="obs")
\end{verbatim}
}

In practice, we may want to use a fitted model for prediction at unobserved
sites. A natural way to achieve prediction in the Bayesian framework
of INLA is to add the prediction points to the data by considering the
associated observations as missing data. To illustrate
\texttt{R-INLA}'s facilities for this approach, we here consider
prediction at instant $t=19$ over a regular spatial grid covering $[0,1]^2$.
Therefore, we first create a separate observation matrix and a stack for the
prediction points with missing data, and we will then use
\texttt{R-INLA}'s \emph{join}-mechanism that allows regrouping several
 groups of predictors $\bm \eta_k$ through their observation matrices
 $\bm A_k$. The \texttt{inla.stack.join(stack1,stack2,...)}-function creates a
single model corresponding to a structure $(\bm A_1',\bm A_2',...)'\bm
x=(\bm\eta_1',\bm\eta_2',...)'$, where information relative to $\bm A_k\bm
x=\bm \eta_k$ is regrouped in  \texttt{stackk} for $k=1,2,\ldots$
{\footnotesize
\begin{verbatim}
n.grid=51
xgrid=0:(n.grid-1)/(n.grid-1)
grid.pred=as.matrix(expand.grid(xgrid,xgrid))
A.pred=inla.spde.make.A(mesh,loc=grid.pred,index=1:nrow(grid.pred),
   group=rep(t.pred,nrow(grid.pred)),n.spde=spde$n.spde,n.group=n.repl)
stack.pred=inla.stack(data=list(y=rep(NA, n.grid^2)),A=list(A.pred,1),
   effects=list(idx.spatial,data.frame(intercept=1,
      covar1=rep(covar1[t.pred],n.grid^2),
         covar2=rep(covar2[t.pred],n.grid^2))),tag="pred")
stack=inla.stack(stack.obs,stack.pred)
\end{verbatim}
}
We now run the estimation with the \texttt{inla(...)} function. Its
syntax ressembles that of \texttt{R}'s \texttt{glm(...)}-function for
generalized linear models, although with a variety of extensions and additional
arguments. 
We need a model formula given in the usual \texttt{R} notation. For
better handling of the intercept term, it is often preferable to
make it appear explicity (\texttt{form=y~-1+intercept+...}), such that
it can later be directly included into the latent space(-time)
model. Fixed effects (\emph{i.e.}, the intercept and covariates whose
linear regression coefficients
are estimated) are added to the formula in the usual way, whereas random effects are added with the
\texttt{f(...)} function. In our model, the approximate SPDE solution
is a random effect. The first argument of \texttt{f} is the name of
the "covariate" associated to the random effect. Having created an
index with name \texttt{spatial} beforehand, we now have a covariate \texttt{spatial}
in the data set that indicates the triangulation node index of the spatial SPDE model
 (repeated $60$ times since the spatial model is duplicated for each
observation time). For prior SPDE models, we further specify the argument
\texttt{model=spde} in \texttt{f}, where \texttt{spde} is the
\texttt{R} object already created for the SPDE prior model. The SPDE model is of purely
spatial nature whereas we have observations in space and time, such
that we can use the \emph{group}-functionality of \texttt{R-INLA} to
define the type of dependence between the $60$ groups of spatially
indexed Gaussian variables. Corresponding to the simulated model, we
here use an AR(1)-group model that models site-wise first-order
autoregression of variables over time. Since we have created the index
\texttt{spatial}, we can specify the argument \texttt{group=spatial.group} to
indicate group membership of the covariates, and we have to set
\texttt{control.group=list(model="ar1")} for the AR(1)-model between
groups. 

\begin{figure}
\centering
\includegraphics[width=12cm]{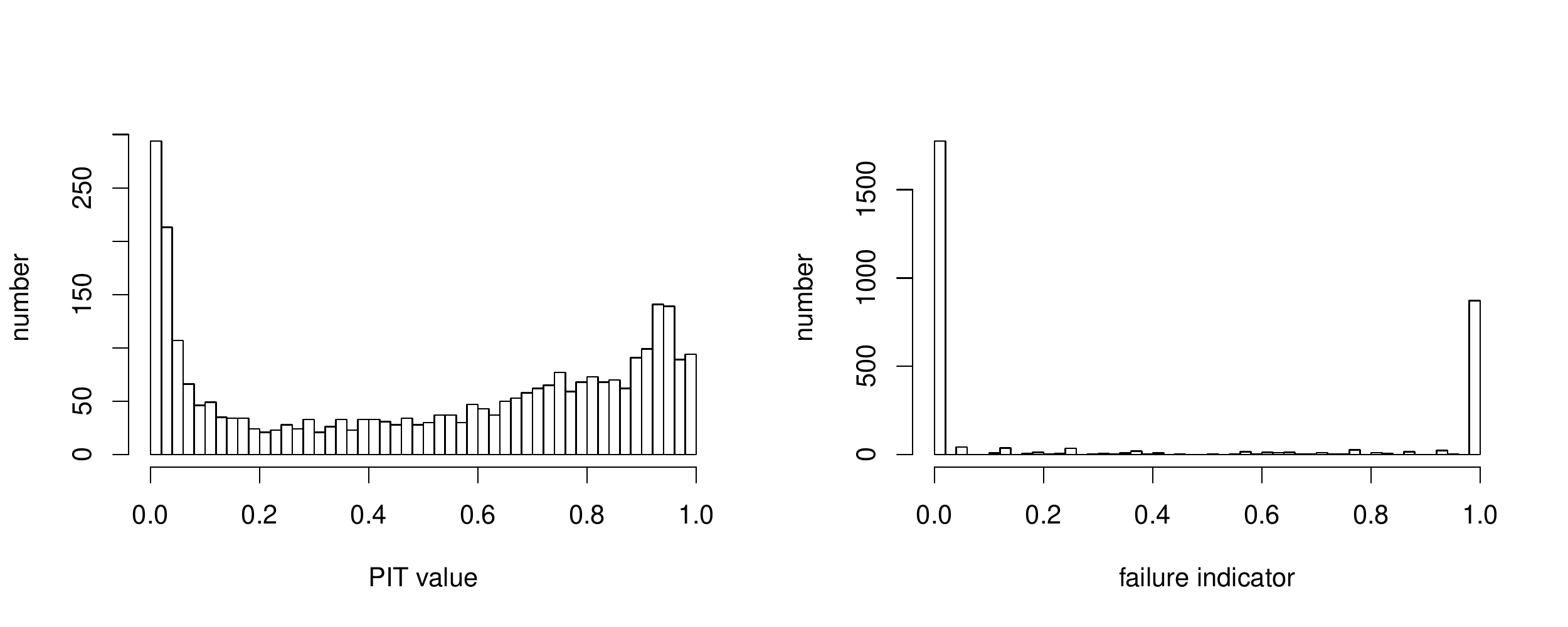}
\caption{Left: histogram of Internally cross-validated cdf values for $\pi(y_i\mid \bm
y_{-i})$. Right: histogram of the failure indicator of trustworthiness
of calculated cdf values. }
\label{fig:pit}
\end{figure}

Data must be passed to the fitting function \texttt{inla(...)} as a
\texttt{data.frame} or \texttt{list}, and the
\texttt{inla.stack.data}-function allows convenient extraction of the preprocessed
\texttt{data} object from the stack. Further control arguments to \texttt{inla(...)} can be
specified explicitly through \texttt{R}'s usual \texttt{control.$\star$=list(...)}
syntax, which allows overriding the default control arguments. Here, $\star$ should be replaced by one of \texttt{inla} (for
controlling INLA-related details like the reordering scheme used for
making the precision matrix as diagonal as possible), \texttt{compute}
(for specifying which quantities should be calculated, \emph{e.g.}
 goodness-of-fit and model selection criteria),
\texttt{predictor} (for specifying the observation matrix $\bm A$ if there
is one, and for indicating which posterior quantities should be calculated for
the predictor vector $\bm \eta$), \texttt{family} (for modifying the
default prior of likelihood hyperparameters), amongst others. 
The choice of prior distributions is often not a straightforward exercise in Bayesian statistics. \texttt{R-INLA} proposes default choices, as for instance non informative priors for fixed effects, but the user can override the default settings by using the \texttt{hyper=list(...)} syntax in the \texttt{control.family} list (for hyper parameters related to the likelihood family) or in the \texttt{f(...)} function when generating random effects; for fixed effects, the \texttt{mean} and \texttt{precision} elements of the \texttt{control.fixed} list allow modifying the prior.   
In the
following, we here fix the METIS-reordering strategy in
\texttt{control.inla} to avoid the higher memory requirements of the standard reordering scheme  (which were too high for the machine with $16$GB of memory used for fitting, leading to a "bus error").  In \texttt{control.predictor}, we pass the observation matrix
$\bm A$ that can be extracted from the stack via \texttt{inla.stack.A(stack)}, and we advise the program
to calculate (discretized) posterior densities for the predictor variables
$\eta_i$  (\texttt{compute=T}). Moreover, for a correct prediction of the \texttt{NA} values, we must tell \texttt{inla} to use the link function from the first likelihood family in \texttt{control.family} (in our case, there is only one) by indicating \texttt{link=1}. By default, \texttt{R-INLA} would have assumed an identity link for \texttt{NA} values.  In \texttt{control.compute}, we here demand the calculation 
of \textsc{CPO}-values (cross-validated predictive measures, see
\citet{Held.al.2010} for a comparison of Markov chain Monte Carlo and INLA), the
marginal likelihood $\pi(\bm y)$, the Deviance Information Criterion
(\textsc{DIC}) and the
Watanabe--Akaike Information Criterion (\textsc{WAIC}, \citet{Watanabe.2010,Gelman.al.2014}), where the
latter can be considered as a Bayesian variant of the common
\textsc{AIC}. 
The \textsc{CPO}-related values are the density and cdf of the posterior
distribution $\pi(y_i\mid
\bm y_{-i})$, evaluated at the observed $y_i$. With INLA, these
cross-validation quantities can be calculated quickly without explicitly
reestimating the model, see \citet{Rue.Martino.Chopin.2009} for details. However, certain
theoretical assumptions might be violated such that some or all of these
CPO-related values are not trustworthy, which is then indicated in
the \texttt{inla}-output in \texttt{fit\$cpo\$failure}. In such a case, the 
\texttt{inla.cpo(...)}-function can be used for  ``manual'' reestimation of
the cross-validated model for the concerned data points $y_i$. 
We now construct the \texttt{inla(...)}-call. For later use,
we here also store the data and parameters of this first model in an object
\texttt{mod1}:
{\footnotesize
\begin{verbatim}
data=inla.stack.data(stack,spde=spde)
form=y~-1+intercept+covar1+covar2+f(spatial,model=spde,
   group=spatial.group,control.group=list(model="ar1"))
cc=list(cpo=T,dic=T,mlik=T,waic=T)
cp=list(A=inla.stack.A(stack),compute=T,link=1)
ci=list(reordering="metis")
mod1=list(stack=stack,data=data,A.pred=A.pred,A.obs=A.obs,
   idx.spatial=idx.spatial,spde=spde,form=form,cp=cp,ci=ci,cc=cc)
fit=inla(form,family="poisson",data=data,control.compute=cc,
   control.predictor=cp,control.inla=ci)
\end{verbatim}
}
Here we have used the default prior for $a$, which is defined as a Gaussian prior with initial value $2$, mean $0$ and precision $0.15$ on $\log((1+a)/(1-a))$. We could have modified it by specifying the \texttt{hyper} argument in \texttt{control.group}; for example, \texttt{control.group =list(model="ar1", hyper=list(rho=list(prior="normal", initial=0, param=c(0,25)))} would keep the Gaussian prior and set a high precision of $25$, therefore leading to an informative prior concentrating strongly around the value $a=0$ resulting in temporal independence.
Since our likelihood is not Gaussian (but Poisson) and
since the latent model is quite complex, the \texttt{inla} run takes some
time (around $50$ minutes on a state-of-the art 4 core machine), and
memory requirements are rather high. Notice that the standard reordering scheme for the precision matrix could lead to a reduced computation time. 
We remark that \texttt{inla(...)} has a \texttt{num.threads} argument 
which allows fixing the maximum number of computation threads used by
\texttt{R-INLA}. By default, \texttt{R-INLA} takes control over all available cores
of the machine for parallel execution, which can lead to problems in
terms of too high memory requirements since each thread occupies a certain
amount of memory. 
As it is usual in \texttt{R}, we can now call \texttt{summary(fit)} to obtain
principal results of the fit. Part of  its output is as follows:
{\footnotesize
\verbatimfont{\itshape}
\begin{verbatim}
[...]
Time used:
 Pre-processing    Running inla Post-processing           Total 
         1.0956       2766.6987          1.0865       2768.8808 
Fixed effects:
             mean     sd 0.025quant 0.5quant 0.975quant    mode kld
intercept -0.9094 0.1451    -1.1959  -0.9091    -0.6251 -0.9084   0
covar1     0.7827 0.2352     0.3193   0.7829     1.2444  0.7833   0
covar2     0.5092 0.0262     0.4579   0.5091     0.5610  0.5089   0
Random effects:
Name	  Model
 spatial   SPDE2 model 
Model hyperparameters:
                        mean     sd 0.025quant 0.5quant 0.975quant    mode
Theta1 for spatial   -3.5143 0.0670    -3.6459  -3.5142    -3.3828 -3.5139
Theta2 for spatial    2.2750 0.0696     2.1385   2.2750     2.4119  2.2750
GroupRho for spatial  0.4892 0.0325     0.4256   0.4889     0.5532  0.4875

Expected number of effective parameters(std dev): 1020.18(24.28)
Number of equivalent replicates : 2.941 
Deviance Information Criterion (DIC) ...: 8544.98
Effective number of parameters .........: 979.41
Watanabe-Akaike information criterion (WAIC) ...: 8522.71
Effective number of parameters .................: 745.45
Marginal log-Likelihood:  -4752.98 
[...]
\end{verbatim}
}
We see that the posterior means of the autoregression coefficient (\texttt{GroupRho}) and
of covariate coefficients of \texttt{covar1} and \texttt{covar2} are not far from the
actually simulated values, and the true values of those parameters lie clearly inside the $95\%$ credible intervals.
The object \texttt{fit} is of
\texttt{list} type; its various slots contain a multitude of information. For
our model, we could be interested in a better interpretable
representation of the hyperparameter estimates of the spatial SPDE model in terms of effective range and variance. In
the following, \texttt{inla.spde.result(...)} extracts the fitting result
for the spatial index \texttt{spatial}  associated to the SPDE. Then, for instance, \texttt{inla.qmarginal}
allows extracting posterior marginal quantiles, and \texttt{inla.emarginal(FUN,
  ...)} calculates posterior marginal expectations of $\mathrm{FUN}(X)$, where
$\mathrm{FUN}$ is a function and 
$X$ is the posterior marginal distribution (note that expectations are
particular since
$\mathbb{E}\mathrm{FUN}(X)\not=\mathrm{FUN}(\mathbb{E}X)$).
{\footnotesize
\begin{verbatim}
result.spatial=inla.spde.result(fit,"spatial",spde)
inla.emarginal(identity,result.spatial$marginals.range.nominal[[1]])
\end{verbatim}
{\verbatimfont{\itshape}
\begin{verbatim}
[1] 0.2914313
\end{verbatim}
}
\begin{verbatim}
inla.qmarginal(c(0.025,0.25,0.5,0.75,0.975),
   result.spatial$marginals.range.nominal[[1]])
\end{verbatim}
{\footnotesize
{\verbatimfont{\itshape}
\begin{verbatim}
[1] 0.2539053 0.2773829 0.2906817 0.3046177 0.3327678
\end{verbatim}
}
}
{\footnotesize
\begin{verbatim}
inla.emarginal(identity,result.spatial$marginals.variance.nominal[[1]]) 
\end{verbatim}
{\verbatimfont{\itshape}
\begin{verbatim}
[1] 0.9515285
\end{verbatim}
}
\begin{verbatim}
inla.qmarginal(c(0.025,0.25,0.5,0.75,0.975),
   result.spatial$marginals.variance.nominal[[1]])
\end{verbatim}
{\verbatimfont{\itshape}
\begin{verbatim}
[1] 0.8368486 0.9083829 0.9490067 0.9918896 1.0788070
\end{verbatim}
}
}
}
Summary statistics for any marginal distribution can further be obtained through the function \texttt{inla.zmarginal(...)}.
We also plot a  histogram of the cross-validated $\pi(y_i\mid \bm
y_{-i})$ cdf values, and give a histogram of the
\texttt{fit\$cpo\$failure} values (between $0$ and $1$), where values
far from $0$ indicate a violation of internal assumptions in the
calculation of cdf values (see Figure \ref{fig:pit} for the resulting
displays):
{\footnotesize
\begin{verbatim}
hist(fit$cpo$pit,breaks=50,main="",xlab="PIT value",ylab="number")
hist(fit$cpo$failure,breaks=50,main="",xlab="failure indicator",ylab="number")
\end{verbatim}
}
Cdf values are not very far from being uniform, which indicates that there are
no strong systematic biases in posterior predictions made with the
model. A certain proportion of the failure indicator values are above
zero and some are even $1$,  meaning that there are some cdf
values that must be interpreted with caution. 

\begin{figure}
\centering
\includegraphics[width=4.25cm]{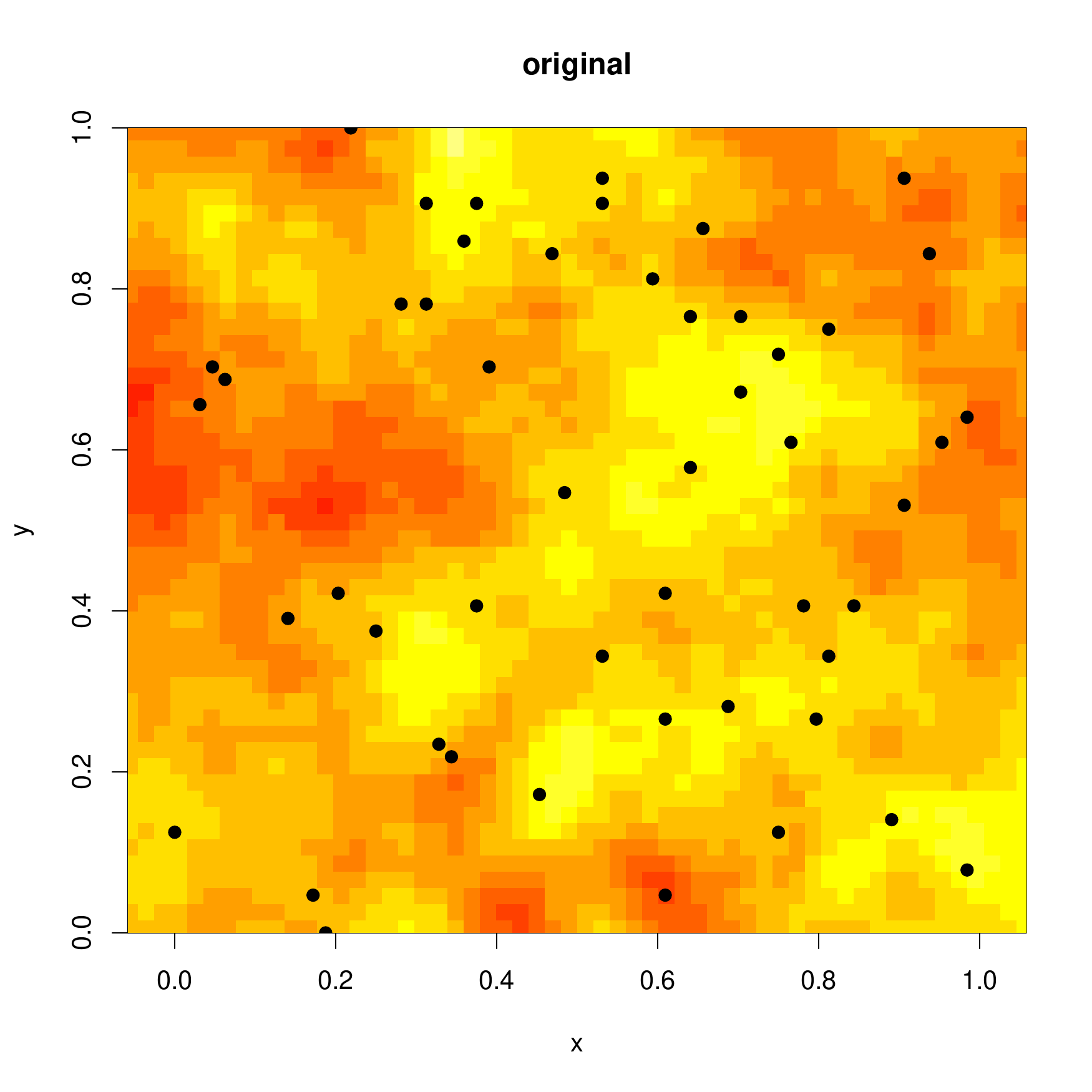}\quad
\includegraphics[width=4.25cm]{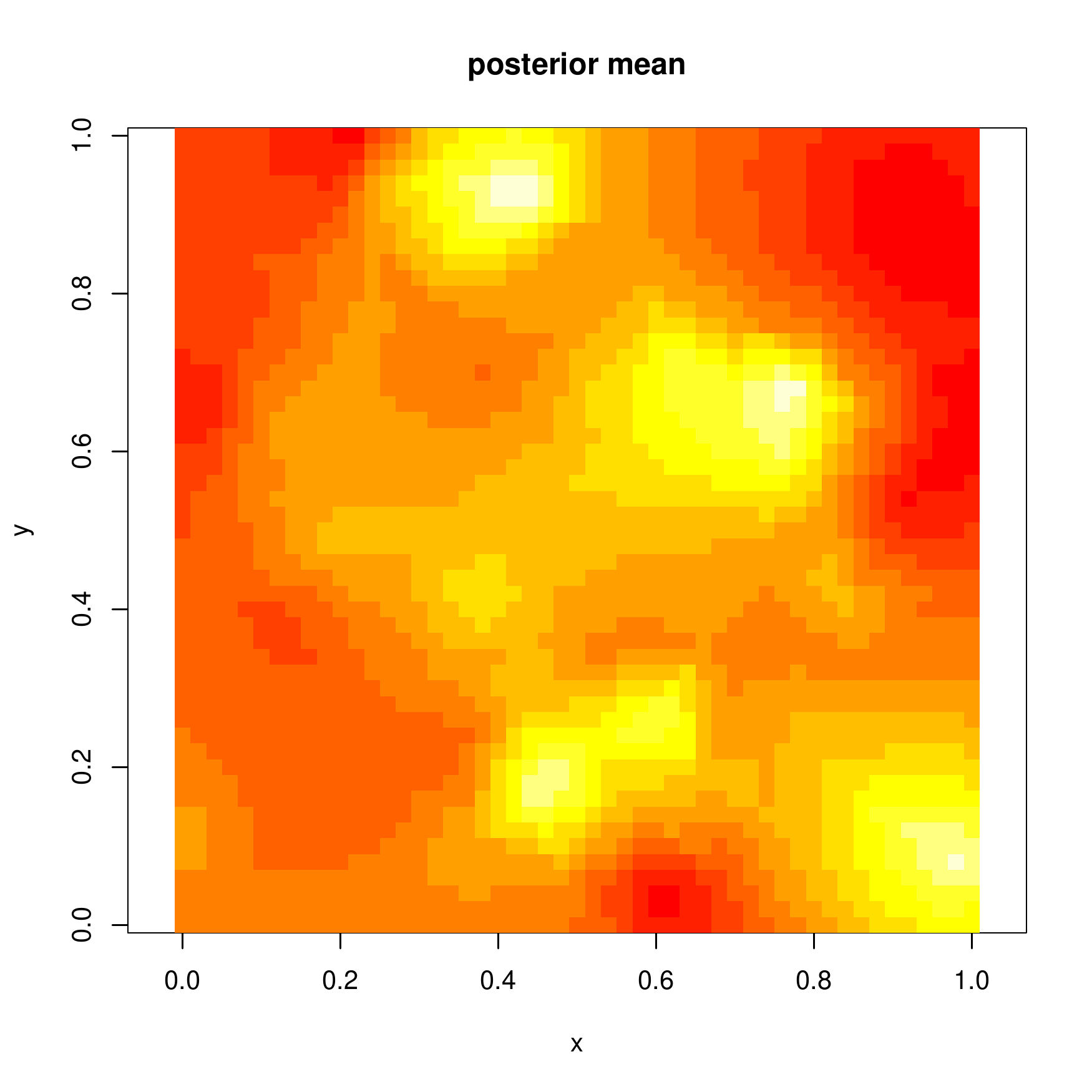}\quad
\includegraphics[width=4.25cm]{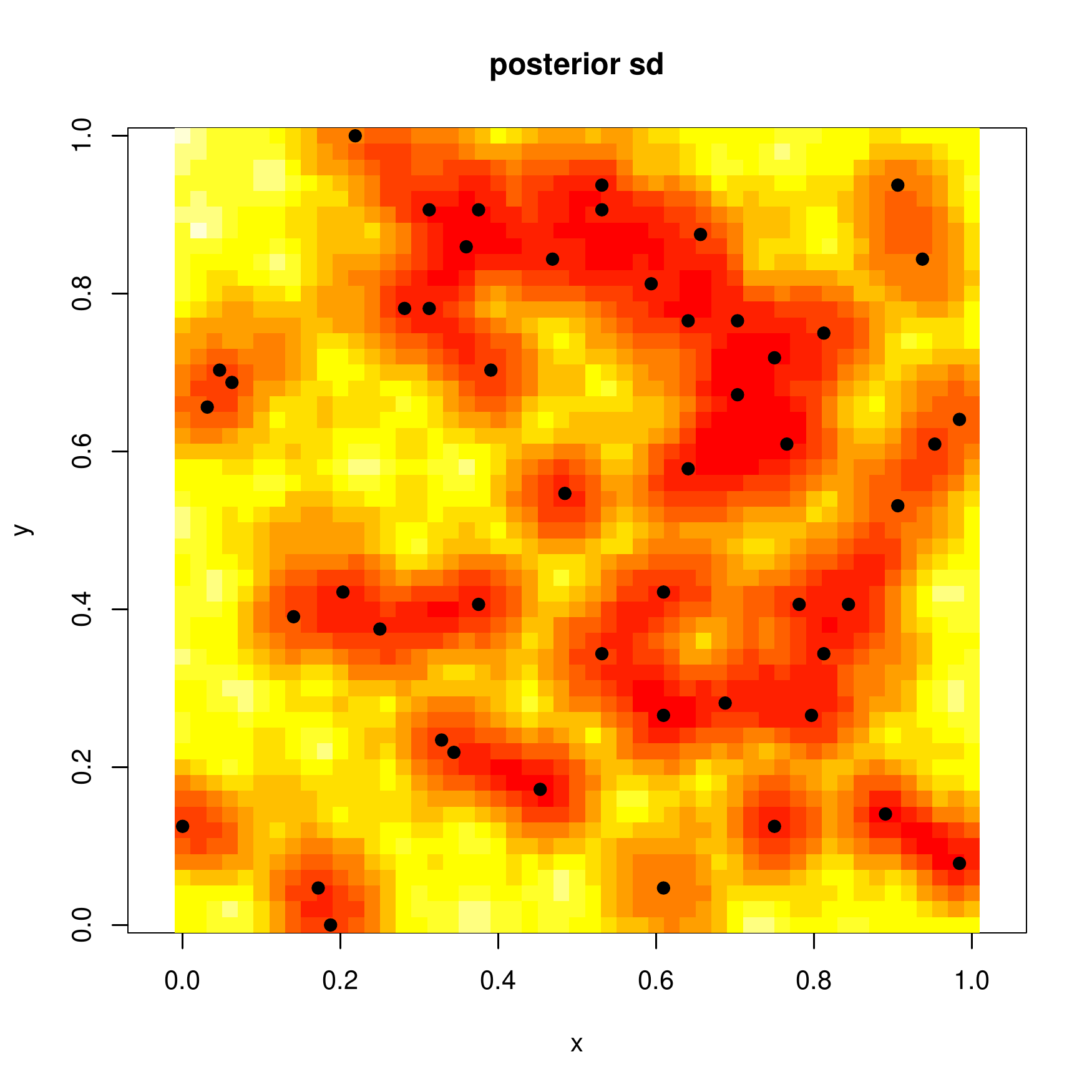}\quad
\caption{Predictions of the latent Gaussian field for $t=19$. Left:
  original simulated data. Middle: posterior mean. Right: posterior
  standard deviation.}
\label{fig:pred}
\end{figure}

Since we have done prediction for $t=19$, we now extract and visualize
the marginal posterior mean and standard deviation over the prediction
grid. To get the index of predicted variables, we can apply the
\texttt{inla.stack.index} function to the \texttt{stack} object by indicating
the prediction sub-stack through the argument \texttt{tag="pred"}. The following code
visualizes the predictions $\hat{\eta}_i$ and the originally simulated
Gaussian values on the prediction
grid. The \texttt{inla.mesh.projector} function allows switching
between the finite element representation and a regular grid by calculating the
Gauss--Markov finite element approximation value to the SPDE for the grid positions
by using the "pyramid" basis functions to interpolate between
triangulation nodes. We use the \texttt{inla.emarginal}-function for
calculating posterior expectations, where the standard deviation is
calculated as $\sqrt{\mathbb{E}X^2-(\mathbb{E}X)^2}$ in the following
code:
{\footnotesize
\begin{verbatim}
idx.pred=inla.stack.index(stack, tag="pred")$data
eta.marginals=fit$marginals.linear.predictor[idx.pred]
eta.mean=unlist(lapply(eta.marginals,function(x) inla.emarginal(identity,x)))
eta.mean=matrix(eta.mean,n.grid,n.grid)
image(x=xgrid,y=xgrid,eta.mean,asp=1,xlab="x",ylab="y",main="posterior mean")
proj=inla.mesh.projector(mesh.sim,xlim=c(0,1),ylim=c(0,1),dims=c(n.grid,n.grid))
image(grid$x,grid$y,inla.mesh.project(proj,field=x[,t.pred]),xlab="x",
   ylab="y",xlim=c(0,1),ylim=c(0,1), asp=1,main="original")
points(sites,pch=19,cex=1)
eta2.mean=unlist(lapply(eta.marginals,function(x) inla.emarginal("^",x,2)))
eta2.mean=matrix(eta2.mean,n.grid,n.grid)
eta.sd=sqrt(eta2.mean-eta.mean^2)
image(x=xgrid,y=xgrid,eta.sd,asp=1,xlab="x",ylab="y",main="posterior sd")
points(sites,pch=19,cex=1)
\end{verbatim}
}
Figure \ref{fig:pred} shows the resulting displays. As expected, uncertainty is lower close to observation sites. The
prediction captures the spatial variation of the actual data, and
a deeper analysis shows that 
the predicted surface is smoother than the original values: since the
Gaussian 
prior on the predictor is centered at $0$, the spatial variation in
posterior predictions is naturally less strong owing to the Bayesian
approach. 

\subsection{Other candidate models}
\label{sec:other}
In Section \ref{sec:fitmod1}, we have used our knowledge about the
simulated data model to construct the Bayesian model that should be the most
appropriate. For comparison and to illustrate \texttt{R-INLA}'s
functionality for other types of models, we here propose to fit some
alternative candidate models. 

For Model 2, we drop the temporal autoregression and consider the
spatial blocks of SPDE variables as independent in the
prior. This necessitates some (slight) adaptations in the code since
there is no more group model with dependence between blocks, but we now
have replicates, \emph{i.e.}, independent blocks.
{\footnotesize
\begin{verbatim}
idx.spatial=inla.spde.make.index("spatial",n.spde=spde$n.spde,n.repl=n.repl)
A.obs=inla.spde.make.A(mesh,loc=sites,index=rep(1:nrow(sites),n.repl),
   repl=rep(1:n.repl,each=nrow(sites)))
stack.obs=inla.stack(data=list(y=y),A=list(A.obs,1),effects=list(idx.spatial,
   data.frame(intercept=1,covar1=rep(covar1,each=n.sites),
      covar2=rep(covar2,each=n.sites))),tag="obs")
A.pred=inla.spde.make.A(mesh,loc=grid.pred,index=1:nrow(grid.pred),
   repl=rep(t.pred,nrow(grid.pred)),n.spde=spde$n.spde,n.repl=n.repl)
stack.pred=inla.stack(data=list(y=NA),A=list(A.pred,1),effects=
   list(idx.spatial,data.frame(intercept=1,covar1=rep(covar1[t.pred],n.grid^2),
      covar2=rep(covar2[t.pred],n.grid^2))),tag="pred")
stack=inla.stack(stack.obs,stack.pred)
data=inla.stack.data(stack,spde=spde)
form=y~-1+intercept+covar1+covar2+f(spatial,model=spde,replicate=spatial.repl)
cp=list(A=inla.stack.A(stack),compute=T)
fit=inla(form,family="poisson",data=data,control.compute=mod1$cc,
   control.predictor=cp,control.inla=mod1$ci)
\end{verbatim}
}
Model 3 focuses on the temporal trend, neglects spatial variation and  supposes that covariates
are not known. Here we use a
temporal  first-order random walk as prior model. We set a rather high
initial prior value for the precision of the random walk innovations
(corresponding to a standard deviation of $0.01$). A sum-to-zero
constraint is added (\texttt{constr=T}) for better identifiability (notice that in the case of the \texttt{rw1} model it is already added by default).
{\footnotesize
\begin{verbatim}
data3=data.frame(y=y,covar1=rep(covar1,each=n.sites))
form=y~f(covar1,model="rw1",hyper=list(prec=list(initial=log(1/.01^2),
   fixed=F)),constr=T) 
cp=list(compute=T)
\end{verbatim}
}
Once the model is fitted, it would be relatively easy to extract information about the posterior distribution of the random trend from the lists \texttt{fit\$summary.linear.predictor} or \texttt{fit\$summary.fitted.values}, which contain $51$ copies of the same posterior information for each time step due to the $51$ sites with data. As an alternative, we here illustrate the powerful \texttt{lincomb}-tool to directly calculate posterior distributions for certain linear combinations of the latent effects, which is very useful whenever  we need precise information on the posterior distribution of some linear combinations of certain latent variables. In our case, the $60$ values of the random trend are represented as  the sum of the intercept (fixed effect) and each of the $n.repl=60$ latent \texttt{rw1} variables (random effect).  The command 
{\footnotesize
	\begin{verbatim}
lc=inla.make.lincombs("(Intercept)"=rep(1,n.repl),covar1=diag(n.repl)) 
	\end{verbatim}
}
defines $60$ linear combinations with structure "intercept value plus $i$th component of the random walk", $i=1,\ldots,60$, where \texttt{"(Intercept)"} refers to the intercept if it has been defined implicitly in the formula without a variable name assigned to it. We now fit the model: 
{\footnotesize
\begin{verbatim}
fit=inla(form,family="poisson",data=data3,lincomb=lc,control.predictor=cp)
\end{verbatim}
}
The \texttt{fit} object will contain a list \texttt{summary.lincomb.derived} providing the requested posterior summaries for the explicitly defined linear combinations of the latent variables.
Another interesting model could be obtained from combining Models 2
and 3, \emph{i.e.}, using a random walk in time and a spatial SPDE
model. Such a model is relatively complex if not overly complicated and its estimation is
computationally demanding, so we do not consider it here. 
In Models 4 and 5, we specify a shape parameter of the Gauss--Markov
Mat\'ern model in the prior  that is different from the simulated
model, using either $\nu=0.5$ (exponential model) or $\nu=0$ (not a
proper Mat\'ern model, but still a valid covariance model). 
{\footnotesize
\begin{verbatim}
nu=0.5 #model 4
nu=0 #model 5
alpha=nu+1
spde=inla.spde2.matern(mesh=mesh,alpha=alpha,B.tau=matrix(c(0,1,0),nrow=1),
   B.kappa=matrix(c(0,0,1),nrow=1))
form=y~-1+intercept+covar1+covar2+f(spatial,model=spde,group=spatial.group,
   control.group=list(model="ar1"))
fit=inla(form,family="poisson",data=mod1$data,control.compute=mod1$cc,
   control.predictor=mod1$cp,control.inla=mod1$ci)
\end{verbatim}
}
Finally, Model 6 uses not the Poisson likelihood but the negative
binomial one that has an additional overdispersion parameter $\theta_{\mathrm{disp}}$, where
the variance of the negative binomial distribution is
$\mu+\mu^2/\theta_{\mathrm{disp}}$ and $\mu$ is its mean. Notice
that the Poisson distribution arises in the limit for
$\theta_{\mathrm{disp}}=\infty$. The internal parametrization of
$\theta_{\mathrm{disp}}$ considers $\log
\theta_{\mathrm{disp}}$ as a hyperparameter. We here fix a relatively
high initial value $\log(10)$ and use an informative log-gamma prior with shape
$10$ and rate $1$, such that the prior expectation of $\theta$ is
$10$ (with prior variance $10$); this yields a prior likelihood model relatively close to the
Poisson one:
{\footnotesize
\begin{verbatim}
cf=list(hyper=list(list(theta=list(initial=log(10),prior="loggamma",
   param=c(10,1)))))
fit=inla(mod1$form,
family="nbinomial",data=mod1$data,control.compute=mod1$cc,
   control.family=cf,control.predictor=mod1$cp, control.inla=mod1$ci)
\end{verbatim}
}
More generally, the use of relatively narrow informative priors may improve the stability of computations in complex models. Recent modifications of \texttt{R-INLA} go towards a more systematic use of the so-called penalized complexity priors \citep{Simpson.al.2014}, designed to shrink the model towards a relatively simple reference model in  a natural way and independently of any reparametrization of prior parameters, where shrinkage towards the reference happens when data do not provide clear evidence of the contrary.
\subsection{Analyzing fitted models}
\label{sec:gof}
\begin{figure}
\centering
\includegraphics[width=6cm]{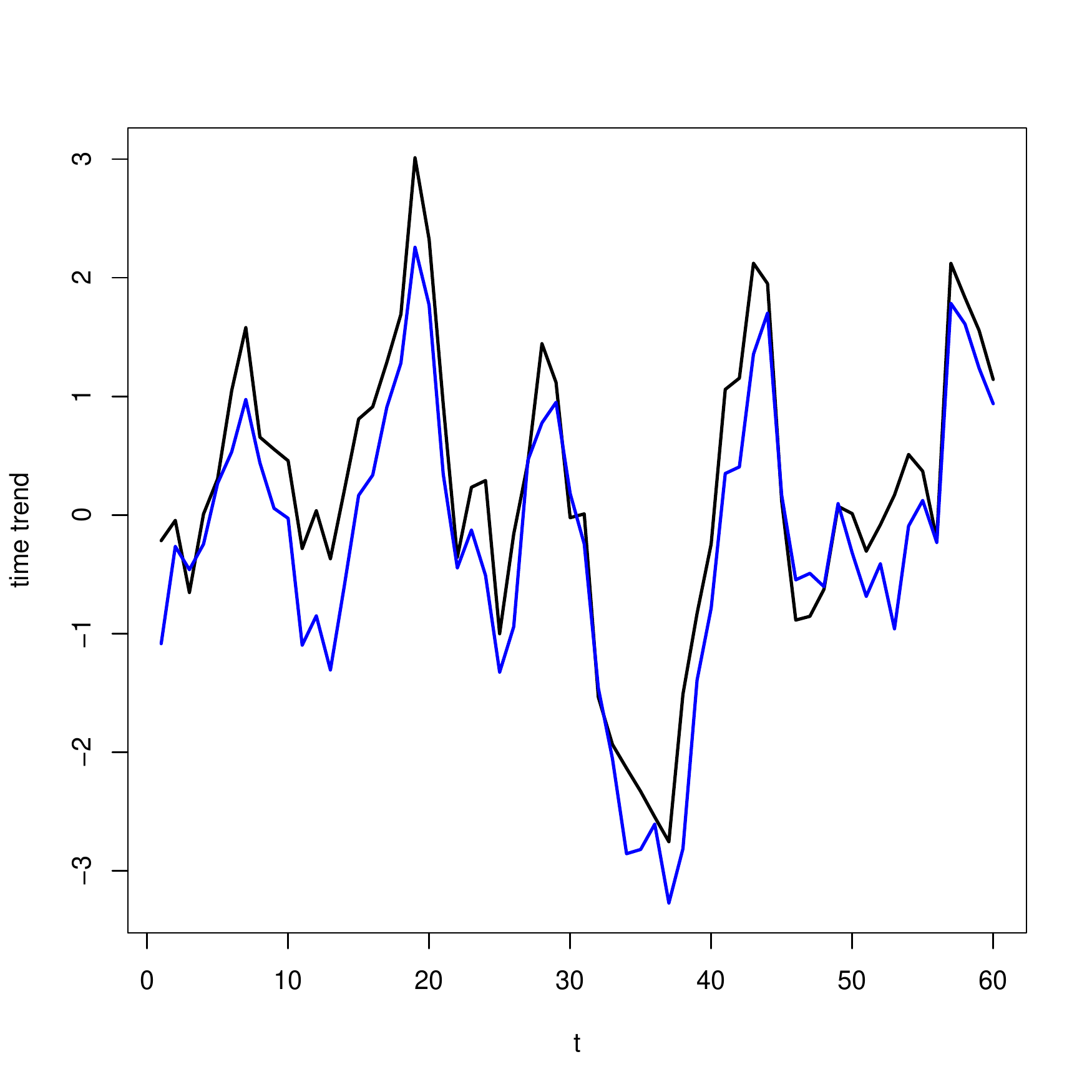}
\caption{Purely temporal random walk model 3. Posterior mean curve (black) and simulated curve (blue).}
\label{fig:temp}
\end{figure}

We now compare the $6$ fitted models. 
For the purely temporal model 3, the following code plots a posterior
mean estimate of the
fitted temporal trend (using the \texttt{lincomb}-feature explained in Section \ref{sec:other}) and the simulated data, see Figure
\ref{fig:temp}:
{\footnotesize
\begin{verbatim}
plot(1:n.repl,fit$summary.lincomb.derived$mean,type="l",xlab="t",
   ylab="time trend",lwd=2)
lines(1:n.repl,xtrend,col="blue", lwd=2)
\end{verbatim}
}
We see that neglecting the spatial variation in data and using a relatively
simple, purely temporal model here permits to reconstruct
quite accurately the simulated temporal trend. 
Finally, we can compare the information criteria  \textsc{DIC} and  \textsc{WAIC},
marginal likelihoods, estimates of spatial range, variance and
 temporal autocorrelation over the candidate
models, see Table \ref{tab:comp}. The marginal likelihood is
relatively close for all space-time models with an explicit spatial
structure and temporal autoregression (Models 1,4,5,6), and has 
its by far lowest value for the purely temporal model 3. Model 1 whose prior
structure is closest to the simulated model turns out best in terms of
\textsc{DIC}, but has slightly higher \textsc{WAIC} values than models
4 and 5 characterized by a different fixed shape parameter $\nu$ in
the SPDE solution, leading to less smooth sample paths in the spatial
prior random field. Notice however that differences in the estimated
\textsc{WAIC} values between models 1,4 and 5
are relatively small such that they should be interpreted with
caution. Model 6 with the
negative binomial likelihood but with the same latent Gaussian prior
model as Model 1  has rather high \textsc{DIC} and \textsc{WAIC}
values, maybe due to approximations and computations that are less
accurate for this fitted model. 
Concerning range and variance parameters of the fitted spatial models,
we find that they are close to simulated values in all cases except
Model 5, where the different prior shape parameter $\nu=0$ in the SPDE
seems having perturbed the calculations of the dependence structure
and the variance. Throughout, the posterior distribution of the autoregression coefficient $a$ (if
estimated) is concentrated around the true value $0.5$. 

\begin{table}
\centering
\begin{tabular}{crrrrrr}
Model & \textsc{DIC} & \textsc{WAIC} & $\mathrm{mlik}$ & $\mathrm{range}$
& $\mathrm{variance}$ & $a$ \\
\hline
1&$8540$&$8520$&$-4750$&$0.29$($0.25$;$0.33$)&$0.95$($0.84$;$1.08$)&$0.49$($0.43$;$0.55$)\\
2&$8660$&$8630$&$-4830$&$0.30$($0.26$;$0.34$)&$0.92$($0.82$;$1.03$)&---\\
3&$9530$&$9270$&$-5860$&---&---&---\\
4&$8550$&$8510$&$-4750$&$0.32$($0.26$;$0.38$)&$1.03$($0.91$;$1.16$)&$0.49$($0.42$;$0.55$)\\
5&$8550$&$8500$&$-4760$&$0.59$($0.43$;$0.79$)&$0.11$($0.09$;$0.14$)&$0.48$($0.42$;$0.55$)\\
6&$8890$&$8810$&$-4750$&$0.34$($0.29$;$0.40$)&$0.82$($0.70$;$0.96$)&$0.53$($0.46$;$0.60$)\\
\end{tabular}
\caption{Comparison of fitted models: \textsc{DIC}, \textsc{WAIC},
  marginal likelihood $\mathrm{mlik}$, spatial range, variance of
  spatial model, autoregression coefficient $a$. All values  are rounded to three
  significant digits and to at most two decimals. Estimates are
  posterior means and $95\%$ credible
  intervals are in parentheses.}
\label{tab:comp}
\end{table}

\section{Discussion}
We have illustrated the theory and practice of Integrated Nested Laplace
Approximation, implemented in the powerful \texttt{R-INLA} software library,  which enables fast and accurate inference of complex
Bayesian models. The dynamic behavior and dependence structure in the
models covered by \texttt{R-INLA} is primarily governed by a latent
Gaussian process for the mean of the univariate data
distribution.   In view of the near endless range of models that are
available, one should perhaps also  sound a note of caution since
users  may be misled to
construct overly complicated  models in practice. 

\texttt{R-INLA} has mechanisms to manage several
different likelihoods in the same model, to use the same latent
variables in different parts of the latent model (``copying'') or to
deal with non-informative missing data. A more detailed explanation of recently added features of \texttt{R-INLA} can be found in \citet{Martins.al.2013} and \citet{Ferkingstad.Rue.2015}, where the latter paper describes an improvement over the default Laplace approximation strategy for difficult modeling cases where the number of latent variables is of the order of the number of observations or where the concavity of the log-likelihood of the data is not strong enough. Although the INLA approach
does not  directly provide the posterior dependence structure between
predictors $\eta_i$ or between the data distributions for
different $y_i$, one could assume a Gaussian copula model with the
Gaussian dependence given by the precision matrix $\bm Q^*$  of the
Gaussian approximation \eqref{eq:gaussappr} to obtain a practically
useful approximation of
the posterior dependence; this approach is implemented through the \texttt{inla.posterior.sample(...)} function in \texttt{R-INLA}. 
New users should
look around at \url{www.r-inla.org},  the main hub for
staying informed about new INLA-related developments, for finding implementation
details of \texttt{R-INLA} and  for  getting advice on specific
problems via its discussion forum. 

The SPDE approach, providing flexible spatial
Gauss--Markov models, is of interest in its own far beyond
the INLA framework where Markovian structures lead to fast
high-dimensional matrix computations. Multivariate extensions \citep{Hu.al.2013} and
certain nested version of SPDEs \citep{Bolin.Lindgren.2011} have
already been proposed in the literature, although they are not yet
available in \texttt{R-INLA}.  Further modeling extensions in
terms of data likelihoods and latent models can be expected in the
near future. At the current stage, certain types of new, user-defined models may be implemented through the \texttt{inla.rgeneric.define(...)} function.  In
particular, the construction of more
complex and realistic Gauss--Markov space-time dependence structures
based on the SPDE approach, capable to
model effects like the nonseparability of space and time would be another big step forward. 
\section{Acknowledgements}
The author is grateful to the editors and two reviewers for many helpful comments and to H\aa vard Rue for a discussion on the Laplace approximation (and for developing \texttt{R-INLA} with the help of  numerous other contributors, of course!). 
\bibliography{biblio}

\begin{thebibliography}{}

\bibitem[Bakka et~al., 2016]{Bakka.al.2016}
Bakka, H., Vanhatalo, J., Illian, J., Simpson, D., and Rue, H. (2016).
\newblock Accounting for physical barriers in species distribution modeling
  with non-stationary spatial random effects.
\newblock {\em arXiv preprint arXiv:1608.03787}.

\bibitem[Bisanzio et~al., 2011]{Bisanzio.al.2011}
Bisanzio, D., Giacobini, M., Bertolotti, L., Mosca, A., Balbo, L., Kitron, U.,
  and Vazquez-Prokopec, G.~M. (2011).
\newblock Spatio-temporal patterns of distribution of {West} {Nile} virus
  vectors in eastern {Piedmont} {Region}, {Italy}.
\newblock {\em Parasit Vectors}, 4:230.

\bibitem[Blangiardo and Cameletti, 2015]{Blangiardo.Cameletti.2015}
Blangiardo, M. and Cameletti, M. (2015).
\newblock {\em Spatial and Spatio-temporal Bayesian Models with R-INLA}.
\newblock John Wiley \& Sons.

\bibitem[Blangiardo et~al., 2013]{Blangiardo.al.2013}
Blangiardo, M., Cameletti, M., Baio, G., and Rue, H. (2013).
\newblock Spatial and spatio-temporal models with {R-INLA}.
\newblock {\em Spatial and spatio-temporal epidemiology}, 7:39--55.

\bibitem[Bolin and Lindgren, 2011]{Bolin.Lindgren.2011}
Bolin, D. and Lindgren, F. (2011).
\newblock Spatial models generated by nested stochastic partial differential
  equations, with an application to global ozone mapping.
\newblock {\em The Annals of Applied Statistics}, pages 523--550.

\bibitem[Cameletti et~al., 2013]{Cameletti.al.2013}
Cameletti, M., Lindgren, F., Simpson, D., and Rue, H. (2013).
\newblock Spatio-temporal modeling of particulate matter concentration through
  the {SPDE} approach.
\newblock {\em AStA Advances in Statistical Analysis}, 97(2):109--131.

\bibitem[Cosandey-Godin et~al., 2014]{CosandeyGodin.al.2014}
Cosandey-Godin, A., Krainski, E.~T., Worm, B., and Flemming, J.~M. (2014).
\newblock Applying {Bayesian} spatiotemporal models to fisheries bycatch in the
  {Canadian} {Arctic}.
\newblock {\em Canadian Journal of Fisheries and Aquatic Sciences},
  72(999):1--12.

\bibitem[Ferkingstad et~al., 2015]{Ferkingstad.Rue.2015}
Ferkingstad, E., Rue, H., et~al. (2015).
\newblock {Improving the INLA approach for approximate Bayesian inference for
  latent Gaussian models}.
\newblock {\em Electronic Journal of Statistics}, 9(2):2706--2731.

\bibitem[Fong et~al., 2010]{Fong.Rue.Wakefield.2009}
Fong, Y., Rue, H., and Wakefield, J. (2010).
\newblock Bayesian inference for generalized linear mixed models.
\newblock {\em Biostatistics}, 11(3):397--412.

\bibitem[Gabriel et~al., 2016]{Gabriel.al.2016}
Gabriel, E., Opitz, T., and Bonneu, F. (2016).
\newblock Detecting and modeling multi-scale space-time structures: the case of
  wildfire occurrences.
\newblock Submitted to Journal de la Soci\'et\'e Fran\c caise de Statistique
  (Special Issue on Space-Time Statistics).

\bibitem[Gelman et~al., 2014]{Gelman.al.2014}
Gelman, A., Hwang, J., and Vehtari, A. (2014).
\newblock Understanding predictive information criteria for bayesian models.
\newblock {\em Statistics and Computing}, 24(6):997--1016.

\bibitem[G{\'o}mez-Rubio et~al., 2015a]{Gomez.al.2015a}
G{\'o}mez-Rubio, V., Bivand, R., and Rue, H. (2015a).
\newblock A new latent class to fit spatial econometrics models with integrated
  nested laplace approximations.
\newblock {\em Procedia Environmental Sciences}, 27:116--118.

\bibitem[G{\'o}mez-Rubio et~al., 2015b]{Gomez.al.2015}
G{\'o}mez-Rubio, V., Cameletti, M., and Finazzi, F. (2015b).
\newblock Analysis of massive marked point patterns with stochastic partial
  differential equations.
\newblock {\em Spatial Statistics}, 14:179--196.

\bibitem[Held et~al., 2010]{Held.al.2010}
Held, L., Schr{\"o}dle, B., and Rue, H. (2010).
\newblock Posterior and cross-validatory predictive checks: a comparison of
  {MCMC} and {INLA}.
\newblock In {\em Statistical modelling and regression structures}, pages
  91--110. Springer.

\bibitem[Hu et~al., 2013]{Hu.al.2013}
Hu, X., Simpson, D., Lindgren, F., and Rue, H. (2013).
\newblock Multivariate {G}aussian random fields using systems of stochastic
  partial differential equations.
\newblock {\em arXiv preprint arXiv:1307.1379}.

\bibitem[Illian et~al., 2012]{Illian.al.2012}
Illian, J.~B., S{\o}rbye, S.~H., Rue, H., et~al. (2012).
\newblock A toolbox for fitting complex spatial point process models using
  integrated nested {L}aplace approximation ({INLA}).
\newblock {\em The Annals of Applied Statistics}, 6(4):1499--1530.

\bibitem[Ingebrigtsen et~al., 2014]{Ingebrigtsen.al.2014}
Ingebrigtsen, R., Lindgren, F., and Steinsland, I. (2014).
\newblock Spatial models with explanatory variables in the dependence
  structure.
\newblock {\em Spatial Statistics}, 8:20--38.

\bibitem[Lindgren and Rue, 2015]{Lindgren.al.2015}
Lindgren, F. and Rue, H. (2015).
\newblock Bayesian spatial modelling with {R-INLA}.
\newblock {\em Journal of Statistical Software}, 63(19).

\bibitem[Lindgren et~al., 2011]{Lindgren.al.2011}
Lindgren, F., Rue, H., and Lindstr{\"o}m, J. (2011).
\newblock An explicit link between {G}aussian fields and {G}aussian {M}arkov
  random fields: the stochastic partial differential equation approach.
\newblock {\em Journal of the Royal Statistical Society: Series B (Statistical
  Methodology)}, 73(4):423--498.

\bibitem[Martino et~al., 2011]{Martino.al.2011}
Martino, S., Akerkar, R., and Rue, H. (2011).
\newblock Approximate {Bayesian} inference for survival models.
\newblock {\em Scandinavian Journal of Statistics}, 38(3):514--528.

\bibitem[Martins et~al., 2013]{Martins.al.2013}
Martins, T.~G., Simpson, D., Lindgren, F., and Rue, H. (2013).
\newblock Bayesian computing with {INLA}: new features.
\newblock {\em Computational Statistics \& Data Analysis}, 67:68--83.

\bibitem[Rue, 2005]{Rue.2005}
Rue, H. (2005).
\newblock Marginal variances for {G}aussian {M}arkov random fields.
\newblock Technical report.
\newblock Norwegian Institute of Science and of Technology, Trondheim.

\bibitem[Rue and Held, 2005]{Rue.Held.2005}
Rue, H. and Held, L. (2005).
\newblock {\em Gaussian {M}arkov random fields: theory and applications}.
\newblock CRC Press.

\bibitem[Rue et~al., 2009]{Rue.Martino.Chopin.2009}
Rue, H., Martino, S., and Chopin, N. (2009).
\newblock Approximate {B}ayesian inference for latent {G}aussian models by
  using integrated nested {L}aplace approximations.
\newblock {\em Journal of the Royal Statistical Society: Series B},
  71(2):319--392.

\bibitem[Rue et~al., 2016]{Rue.al.2016}
Rue, H., Riebler, A., S{\o}rbye, S.~H., Illian, J.~B., Simpson, D.~P., and
  Lindgren, F.~K. (2016).
\newblock Bayesian computing with inla: a review.
\newblock {\em Annual Review of Statistics and Its Application}, 1.

\bibitem[Saumard and Wellner, 2014]{Saumard.al.2014}
Saumard, A. and Wellner, J.~A. (2014).
\newblock Log-concavity and strong log-concavity: a review.
\newblock {\em Statistics surveys}, 8:45.

\bibitem[Schr\"odle and Held, 2011]{Schroedle.al.2011}
Schr\"odle, B. and Held, L. (2011).
\newblock Spatio-temporal disease mapping using {INLA}.
\newblock {\em Environmetrics}, 22(6):725--734.

\bibitem[Schr\"odle et~al., 2012]{Schroedle.al.2012}
Schr\"odle, B., Held, L., and Rue, H. (2012).
\newblock Assessing the impact of a movement network on the spatiotemporal
  spread of infectious diseases.
\newblock {\em Biometrics}, 68(3):736--744.

\bibitem[Serra et~al., 2014]{Serra.al.2014}
Serra, L., Saez, M., Mateu, J., Varga, D., Juan, P., D{\'\i}az-{\'A}valos, C.,
  and Rue, H. (2014).
\newblock Spatio-temporal log-{G}aussian {C}ox processes for modelling wildfire
  occurrence: the case of {C}atalonia, 1994--2008.
\newblock {\em Environmental and Ecological Statistics}, 21(3):531--563.

\bibitem[Simpson et~al., 2014]{Simpson.al.2014}
Simpson, D.~P., Rue, H., Martins, T.~G., Riebler, A., and S{\o}rbye, S.~H.
  (2014).
\newblock Penalising model component complexity: A principled, practical
  approach to constructing priors.
\newblock {\em arXiv preprint arXiv:1403.4630}.

\bibitem[Tierney and Kadane, 1986]{Tierney.Kadane.1986}
Tierney, L. and Kadane, J.~B. (1986).
\newblock Accurate approximations for posterior moments and marginal densities.
\newblock {\em Journal of the American Statistical Association},
  81(393):82--86.

\bibitem[Watanabe, 2010]{Watanabe.2010}
Watanabe, S. (2010).
\newblock Asymptotic equivalence of bayes cross validation and widely
  applicable information criterion in singular learning theory.
\newblock {\em The Journal of Machine Learning Research}, 11:3571--3594.

\end{thebibliography}

\end{document}